\documentclass[iop]{emulateapj}

\usepackage{xcolor}

\usepackage{apjfonts}
\usepackage{graphicx}
\usepackage{epstopdf}
\usepackage{ulem}
\usepackage[utf8]{inputenc}
\usepackage[T1]{fontenc}
\usepackage{url}
\usepackage{mdwlist}
\usepackage{amsmath} 
\usepackage{enumitem}
\usepackage[colorlinks=true, linkcolor = blue, citecolor = blue]{hyperref}

\makeatletter
\def\url@leostyle{%
 \@ifundefined{selectfont}{\def\UrlFont{\sf}}{\def\UrlFont{\small\ttfamily}}}
\makeatother
\begin{document}

\newcommand{\ls}{{_<\atop^{\sim}}}
\newcommand{\gs}{{_>\atop^{\sim}}}
\def \spose#1{\hbox  to 0pt{#1\hss}}  
\def \ls{\mathrel{\spose{\lower 3pt\hbox{$\sim$}}\raise  2.0pt\hbox{$<$}}}
\def \gs{\mathrel{\spose{\lower  3pt\hbox{$\sim$}}\raise 2.0pt\hbox{$>$}}}
\newcommand{\Ha}{\hbox{{\rm H}$\alpha$}}
\newcommand{\Hb}{\hbox{{\rm H}$\beta$}}
\newcommand{\Ovi}{\hbox{{\rm O}\kern 0.1em{\sc vi}}}
\newcommand{\OIII}{\hbox{[{\rm O}\kern 0.1em{\sc iii}]}}
\newcommand{\OII}{\hbox{[{\rm O}\kern 0.1em{\sc ii}]}}
\newcommand{\NII}{\hbox{[{\rm N}\kern 0.1em{\sc ii}]}}
\newcommand{\SII}{\hbox{[{\rm S}\kern 0.1em{\sc ii}]}}
\newcommand{\angstrom}{\textup{\AA}}
\newcommand\ionn[2]{#1$\;${\scshape{#2}}}

\font\btt=rm-lmtk10


\title{Signatures of inflowing gas in red geyser galaxies hosting radio-AGN}

\shorttitle{Neutral gas activities in Red Geysers}
 
\shortauthors{Roy et al.}


\author{Namrata Roy\altaffilmark{1}\dag, 
Kevin Bundy\altaffilmark{1,2}, 
Kate H. R. Rubin\altaffilmark{3},
Kate Rowlands \altaffilmark{4,5}, 
Kyle Westfall \altaffilmark{1,2},
Rogerio Riffel\altaffilmark{6,7},
Dmitry Bizyaev \altaffilmark{8},
David V. Stark \altaffilmark{9},
Rogemar A. Riffel\altaffilmark{10,7},
Ivan Lacerna \altaffilmark{11,12},
Preethi Nair \altaffilmark{13},
Xuanyi Wu \altaffilmark{14},
Niv Drory \altaffilmark{15}}
\altaffiltext{1} {Department of Astronomy and Astrophysics, University of California, 1156 High Street, Santa Cruz, CA 95064}
\altaffiltext{2}{UCO/Lick Observatory, Department of Astronomy and Astrophysics, University of California, 1156 High Street, Santa Cruz, CA 95064}
\altaffiltext{3}{San Diego State University, Department of Astronomy, San Diego, CA 92182, USA}
\altaffiltext{4}{AURA for ESA, Space Telescope Science Institute, Baltimore, MD, USA}
\altaffiltext{5}{Johns Hopkins University, Department of Physics and Astronomy, Baltimore, MD 21218, USA}
\altaffiltext{6}{Departamento de Astronomia, Instituto de F\'\i sica, Universidade Federal do Rio Grande do Sul, CP 15051, 91501-970, Porto Alegre, RS, Brazil}
\altaffiltext{7}{Laborat\'orio Interinstitucional de e-Astronomia - LIneA, Rua Gal. Jos\'e Cristin}
\altaffiltext{8}{	Apache Point Observatory
P.O. Box 59, Sunspot, NM 88349}
\altaffiltext{9}{Haverford College, Department of Physics and Astronomy, 370 Lancaster Ave, Haverford, PA 19041}
\altaffiltext{10}{Departamento de F\'isica, CCNE, Universidade Federal de Santa Maria, 97105-900, Santa Maria, RS, Brazil}
\altaffiltext{11}{Instituto de Astronom\'ia y Ciencias Planetarias, Universidad de Atacama, Copayapu 485, Copiap\'o, Chile}
\altaffiltext{12}{Millennium Institute of Astrophysics, Nuncio Monsenor Sotero Sanz 100, Of. 104, Providencia, Santiago, Chile}
\altaffiltext{13}{	University of Alabama, Department of Physics and Astronomy, Tuscaloosa, AL 35487, USA}
\altaffiltext{14}{Tsinghua University, Department of Astronomy, Beijing 100084, China}
\altaffiltext{15}{McDonald Observatory, The University of Texas at Austin, 1 University Station, Austin, TX 78712, USA}

\altaffiltext{\dag}{naroy@ucsc.edu}


\begin{abstract}

 We study cool neutral gas traced by NaD absorption in 140 local ($\rm z<0.1)$ early-type ``red geyser'' galaxies. These galaxies show unique signatures in spatially-resolved strong-line emission maps that have been interpreted as large-scale active galactic nuclei driven ionized winds. To investigate the possible fuel source for these winds, we examine the abundance and kinematics of cool gas ($\rm T \sim 100-1000 K$) inferred from Na I D absorption in red geysers and matched control samples drawn from SDSS-IV MaNGA.  We find that red geysers host greater amounts of NaD-associated material. Substantial cool gas components are detected in more than $\rm 50 \%$ of red geysers (compared to 25\% of the control sample) going up to 78$\%$ for radio-detected red geysers. Our key result is that cool gas in red geysers is predominantly infalling.  Among our 30 radio-detected red geysers, 86$\%$ show receding NaD absorption velocities (with respect to the systemic velocity) between $\rm 40 - 50~km~s^{-1}$. We verify this result by stacking NaD profiles across each sample which confirms the presence of infalling NaD velocities within red geysers ( $\sim\rm 40~km~s^{-1}$) with no velocity offsets detected in the control samples.  Interpreting our observations as signatures of inflowing cool neutral clouds, we derive an approximate mass inflow rate of $\rm \dot{M}_{in} \sim 0.1 M_{\odot} yr^{-1}$, similar to that expected from minor merging and internal recycling. Some red geysers show much higher rates ($\rm \dot{M}_{in} \sim 5 M_{\odot} yr^{-1}$) that may indicate an ongoing accretion event.


\end{abstract}



\section{Introduction} \label{sec:Introduction}


Low-redshift ($\rm z < 0.1$) integral field unit (IFU) spectroscopy from the Sloan Digital Sky Survey-IV (SDSS-IV) Mapping Nearby Galaxies at Apache Point Observatory (MaNGA) survey \citep{bundy15} has
revealed a population of moderate mass (log $\rm M_{\star}/M_{\odot} \sim 10.5$) passive red galaxies ($ NUV-r > 5$), known as ``red geysers'' \citep{cheung16, roy18}. These galaxies possess unique emission and kinematic properties that may be signatures of galactic scale ($\rm \sim 10~kpc$), centrally-driven outflows \citep{cheung16, roy20}.  The lack of any detectable star-formation \citep[average log SFR ($\rm M_{\odot} / yr ) < -2$, using GALEX+SDSS+WISE from][see Fig.~\ref{fig:sfr}]{salim16}, the association with low luminosity radio-mode active galactic nuclei \citep[AGN,][]{roy18}, and their relatively high occurrence rate on the red sequence \citep[5--10\%, see][]{cheung16}, makes red geysers a candidate for supplying ``maintenance (or radio)-mode'' feedback
by low luminosity AGN.  While such a feedback process has been widely proposed and studied as a way to suppress star formation at late times \citep{binney95, ciotti01, croton06, bower06, faber07, ciotti10, yuan14}, 
direct observational evidence in typical quiescent galaxies has remained elusive. 

Red geysers are characterized by a distinctive bisymmetric pattern in spatially resolved equivalent width (EW) maps of strong  emission lines ($\Ha$, $\NII$, $\OII$).  These features extend to $\sim$ 10 kpc, up to the edge of the MaNGA fiber bundle. The emission line flux distributions are also widespread and significantly elevated ($\sim \rm 5 - 10$~times) along the bi-symmetric feature.  The observed gas is possibly ionized by post asymptotic giant branch (AGB) stars with some contribution from shocks, as evident from a combination of LI(N)ER and Seyfert like line ratios in spatially resolved BPT \citep[Baldwin, Phillips \& Terlevich,][]{baldwin} diagrams in these galaxies \citep{cheung16, roy20}. This enhanced emission roughly aligns with the gas kinematic major axis but is strongly misaligned with the stellar velocity gradient. 


Accreted gas disks in early type galaxies \citep{chen16, lagos14, lagos15, sarzi06, davis13, bryant19, duckworth20} can produce similar misaligned stellar and gas velocity fields so it is important to test the hypothesis that red geysers signal the presence of outflows.  \cite{cheung16} presented detailed dynamical modeling to conclude that the observed ionized gas velocities in the prototypical red geyser were too high to be in gravitationally bound orbits.  More recently, our group used the Keck Echellette Spectrograph and Imager (ESI) to obtain higher spectral resolution observations ($\rm R\sim 8000$ compared to $\rm R\sim 2000$ in MaNGA) of two representative red geysers to find a systematic variation in the asymmetry of the emission line profiles with radius \citep{roy20}.  The observed magnitude and nature of the asymmetry along with an increased velocity dispersion (exceeding $\rm \sim 230~km~s^{-1}$) are consistent with line-of-sight projections through a broad conical outflow.  The alternative scenario of a puffy, rotating gas disk is unable to fit the data.



An additional set of tests involve the mechanism driving the putative wind.
 \cite{riffel19} studied the nuclear region of the prototypical red geyser in \cite{cheung16} with Gemini GMOS (Gemini Multi Object Spectrograph). By comparing the emission line flux distributions and gas kinematics from the inner parts (within 1$''$) with that of the outer regions ($5''$ away from center), they concluded that the observed red geyser show signatures of precession. \cite{roy18}, meanwhile, used the Very large Array (VLA) Faint Images of the Radio Sky at Twenty-Centimeters (FIRST) survey to measure significantly higher ($ > 5 \sigma$) radio continuum flux in stacked red geyser samples and a three times higher radio detection rate compared to control samples. The study concluded that the red geysers host low luminosity radio AGNs ($\rm L_{1.4 GHz} \sim 10^{22} - 10^{23}~W/Hz$) with radiatively inefficient accretion (Eddington scaled accretion rate $\lambda \sim 10^{-4}$).  These AGNs are energetically capable, however, of driving sub-relativistic winds consistent with the MaNGA observations.


What is the fuel source for this AGN activity?  The presence of warm ionized gas suggests the possibility that red geysers also host cooler gas reservoirs in the interstellar medium (ISM). Cool gas in the form of outflows are routinely observed in ultra luminous infrared galaxies, star forming and post-starburst galaxies \citep{cazzoli16, rubin12, rubin14, weiner09, alatalo16, martin05, chen10, yesuf17, borsani18, coil11, rupke18, veilleux20}.  Indeed, 
\citet{cheung16} studied signatures of cool material traced by NaD absorption in their prototypical red geyser.  With a mass estimated at $\rm M_{cool} \sim 10^{8}~M_{\odot}$, the NaD component occupied one side of the galaxy, kinematically and spatially distinct from the ionized gas and apparently inflowing at $\sim \rm 60~km~s^{-1}$ towards the center.  An idealized merger simulation run by these authors suggested that the cool gas was being accreted from a companion galaxy.


Detecting gaseous inflows across the galaxy population in general has garnered significant attention, but definitive inflow signatures have been hard to detect \citep[although see][]{sato09, krug10, rubin12, putman12, zheng17,  sarzi16}. The advent of IFU surveys is revolutionizing this field with the measurement of spatially resolved absorption line kinematics in large samples of galaxies \citep[][Rowlands et al in prep.]{rupke21}.  Here, the kinematics of cool gas are computed by estimating the doppler shifts of the suitable lines with respect to the systemic velocity after carefully accounting for the stellar continuum \citep[down-the-barrel technique,][]{borsani18}. This method involves probing the gas lying in front of the galaxy, with the continuum arising due to the illumination by the background stellar light of the galaxy. Redshifted absorption indicates inflows, i.e., the motion of the gas towards the galaxy (or away from the observer along the line-of-sight). Similarly, a blueshifted component suggests outflows.  Different tracers are available in rest-frame UV or
optical wavelengths \citep[e.g., Fe II $\rm \lambda \lambda$2586,2600, Mg II $\rm \lambda \lambda$2796,2803, Na I D $\rm \lambda \lambda$5891,5897; ][]{chen10, rubin12,rubin14, rupke18, veilleux20}. In this work we use the resonant Na I absorption doublet at 5891 \AA~and 5897 \AA~(referred to as NaD) which traces cool (T $\sim$ 100--1000 K), metal-enriched gas.

An important aspect of measuring the NaD-associated gas kinematics correctly is subtracting the stellar continuum with an appropriate accuracy. This is especially important in early type galaxies, since the evolved K-M type stars present throughout the galaxy can contribute a significant fraction to sodium absorption in the spectrum, thus contaminating the signal from the ISM. There exists several stellar population model libraries (theoretical and empirical) which can be used to estimate the stellar component of NaD absorption generally within a variation of $\sim \rm 10-30 \%$. This difference in the continuum estimate due to the specific choice of the stellar model can affect the estimated properties of the ISM gas. Spatially resolved spectra from MaNGA can aid this process significantly by revealing irregularities in the 2D morphology of the NaD absorption. When these spatial irregularities have no correspondence with the stellar velocity and the surface brightness distribution, we can be more confident that they are associated with the ISM. 


 The goal of this work is to address how frequent the cool gas as traced by NaD is within the red geyser sample and whether these cooler gas reservoirs are associated with inflows towards the center, similar to the prototypical red geyser, 
or with the outflowing wind triggered by the AGN. We will also study if there exists any possible correlation between radio properties and the presence of cool gas, by conducting our analyses on samples split by radio detection.  We analyze the 
spatially resolved, globally integrated and stacked kinematics of NaD from the red geyser sample and compare them with a matched control sample. We find that the red geysers, specially those which are detected in FIRST, have higher NaD EW on average and 78$\%$ of the sample show detectable ISM gas over a projected spatial extent $> 5~\rm kpc$. This fraction is many times higher compared to the radio-detected control (41$\%$) and non radio detected control (25$\%$) galaxies. The kinematics is observed to be redshifted on average ($\sim \rm 40 - 50~km~s^{-1}$) in about $\sim \rm 86\%$ of radio-detected red geysers, implying that most of the gas is inflowing into the galaxy possibly fuelling the central active black hole. 

\S\ref{sec:data} give a brief overview of the optical data from MaNGA that we have used in this work and in \S\ref{sample} we describe the sample selection process of the red geyser and control galaxy samples from the MaNGA quiescent population. The technical details of the stellar continuum modeling, equivalent width estimation, absorption line fitting and kinematics calculation are narrated in sub-sections \S\ref{stellar}, \S\ref{ew} and \S\ref{fitting}. The results of the analyses are described in \S\ref{results} and the implications and concluding remarks are discussed in \S\ref{discussion} and \S\ref{conclusion}. 
Throughout this paper, we assume a flat cosmological model with $H_{0} = 70$ km s$^{-1}$ Mpc$^{-1}$, $\Omega_{m} = 0.30$, and  $\Omega_{\Lambda} =0.70$, and all magnitudes are given in the AB magnitude system. 

\section{Data Acquisition and Sample Definitions } 

\subsection { The MaNGA survey}  \label{sec:data}

We use data primarily from the recently completed SDSS-IV MaNGA survey \citep{blanton17, bundy15, drory15, law15, yan16, sdss16}. MaNGA is an integral field spectroscopic survey that provides spatially resolved spectroscopy for nearby galaxies ($\rm z\sim0.03$) with an effective spatial resolution of $2.5''$ (full width at half-maximum; FWHM). The MaNGA survey uses the SDSS 2.5 meter telescope in spectroscopic mode \citep{gunn06} and the two dual-channel BOSS spectrographs \citep{smee13} that provide continuous wavelength coverage from the near-UV to the near-IR: $\rm3,600-10,000$ \AA. The spectral resolution varies from $\rm R\sim1400$ at 4000~\AA~ to $\rm R\sim2600$ at 9000~\AA. An $r$-band signal-to-noise $(S/N)$ of $\rm 4-8$~\AA$^{-1}$ is achieved in the outskirts (i.e., $\rm1-2~R_{e}$) of target galaxies with an integration time of approximately 3-hr. 
MaNGA has observed more than 10,000 galaxies with $\rm \log~(M_*/ M_{\odot})\gs9$ across $\sim$ 2700 deg$^{2}$ over its 6~yr duration. In order to balance radial coverge versus spatial resolution, MaNGA observes two thirds of its galaxy sample to $\sim$ 1.5~R$_e$ and one third to 2.5~R$_e$. The MaNGA target selection is described in detail in \cite{wake17}.

The raw data are processed with the MaNGA Data Reduction Pipeline (DRP) \citep{law16}. An individual row-by-row algorithm is used to extract the fiber flux and derive inverse variance spectra from each exposure, which are then wavelength calibrated, flat-fielded and sky subtracted. We use the MaNGA sample and data products drawn from the MaNGA Product Launch-9 (MPL-9) and Data Release 16 \citep[DR16,][]{ahumada20}. We use spectral measurements and other analyses carried out by  MaNGA Data Analysis Pipeline (DAP), specifically version 2.3.0. 
The data we use here are based on DAP analysis of each spaxel in the MaNGA datacubes.  
The DAP first fits the stellar continuum of each spaxel to determine the stellar kinematics using the Penalised Pixel-fitting algorithm {\tt pPXF} \citep{cappellari04, cappellari17}  and templates based on the MILES stellar library \citep{MILES}.  The templates are a hierarchically clustered distillation of the full MILES stellar library into 49 templates.  This small set of templates provide statistically equivalent
fits to those that use the full library of 985 spectra in the MILES stellar library.  The emission-line regions are masked during this fit.
The DAP then subtracts the result of the stellar continuum modeling to provide a (nearly) continuum-free spectrum that is used to fit the nebular emission lines.  This version of the DAP treated each line independently, fitting each for its flux, Doppler shift, and width, assuming a Gaussian profile shape.  The final output from the DAP are gas and stellar kinematics, emission line properties and stellar absorption indices. All the spatially resolved 2D maps shown in the paper are outputs from the DAP with hybrid binning scheme. An overview of the DAP used for DR15 and its products is described by \cite{westfall19}, and assessments of its emission-line fitting approach is described by \cite{belfiore19}.

We use ancillary data drawn from the NASA-Sloan Atlas\footnote{\href{http://www.nsatlas.org}{http://www.nsatlas.org}} (NSA) catalog which reanalyzes images and derives morphological parameters for local galaxies observed in Sloan Digital Sky Survey imaging. It compiles spectroscopic redshifts, UV photometry (from GALEX; \citealt{martin05}), stellar masses, and structural parameters. We have specifically used spectroscopic redshifts and stellar masses from the NSA catalog.

\subsection{Sample definitions} \label{sample}

\begin{figure}[h!!!] 
\centering
\graphicspath{{./plots/}}
\includegraphics[width = 0.5\textwidth]{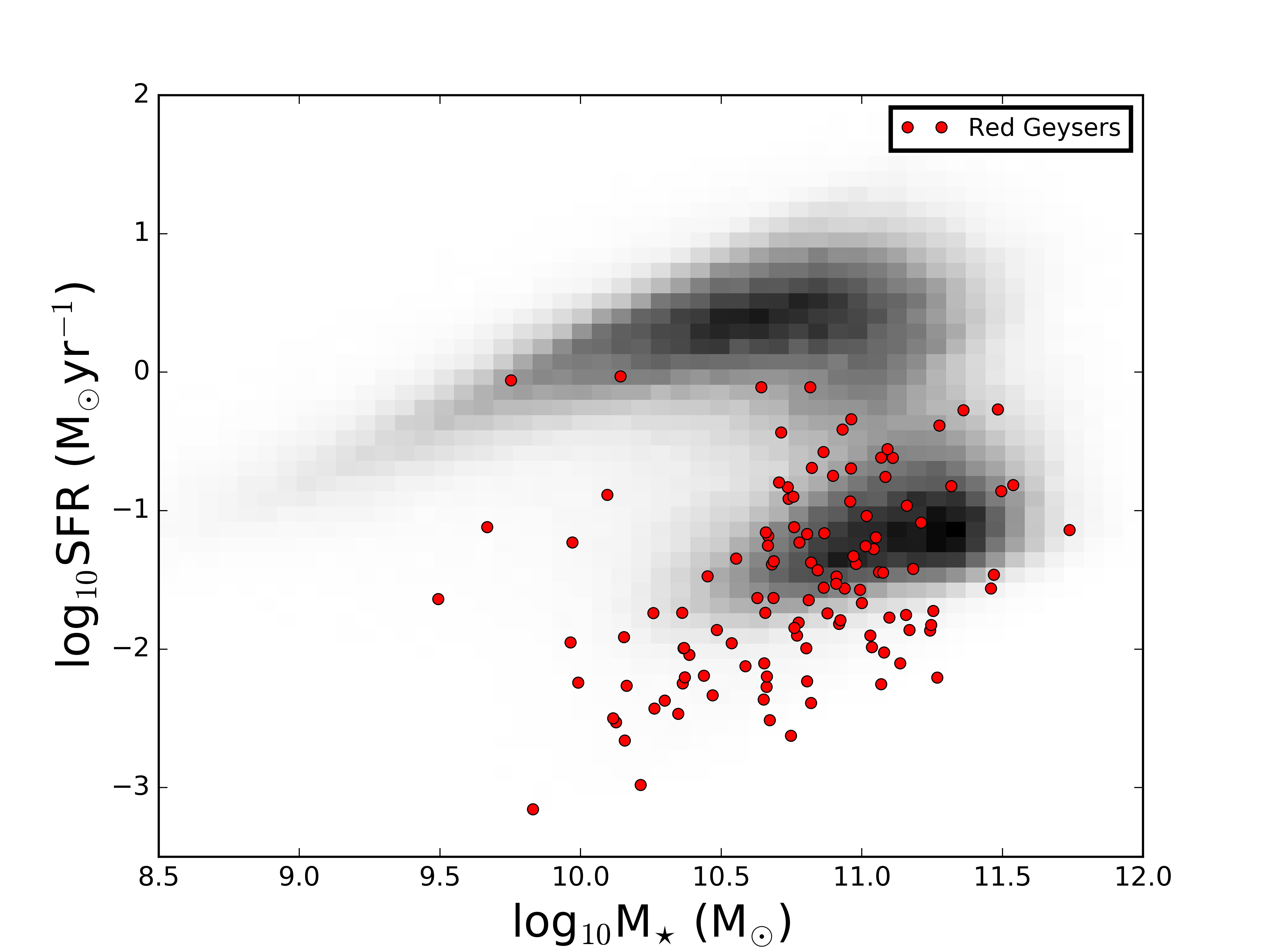}
\caption{The figure shows the log SFR vs log M$_\star$ as obtained from GALEX+SDSS+WISE catalog of \cite{salim16}. The gray 2D histogram shows all the MaNGA galaxies
in the catalog with 0.01 < z < 0.1. The red circles signifies red geyser galaxies. Most of the galaxies in our chosen geyser
sample have a low log SFR value, with an average of $\sim 0.01 \rm M_\odot yr^{-1}$.
\label{fig:sfr}}
\end{figure}

\begin{figure*}[h!!!] 
\centering
\graphicspath{{./plots/}}
\includegraphics[width = \textwidth]{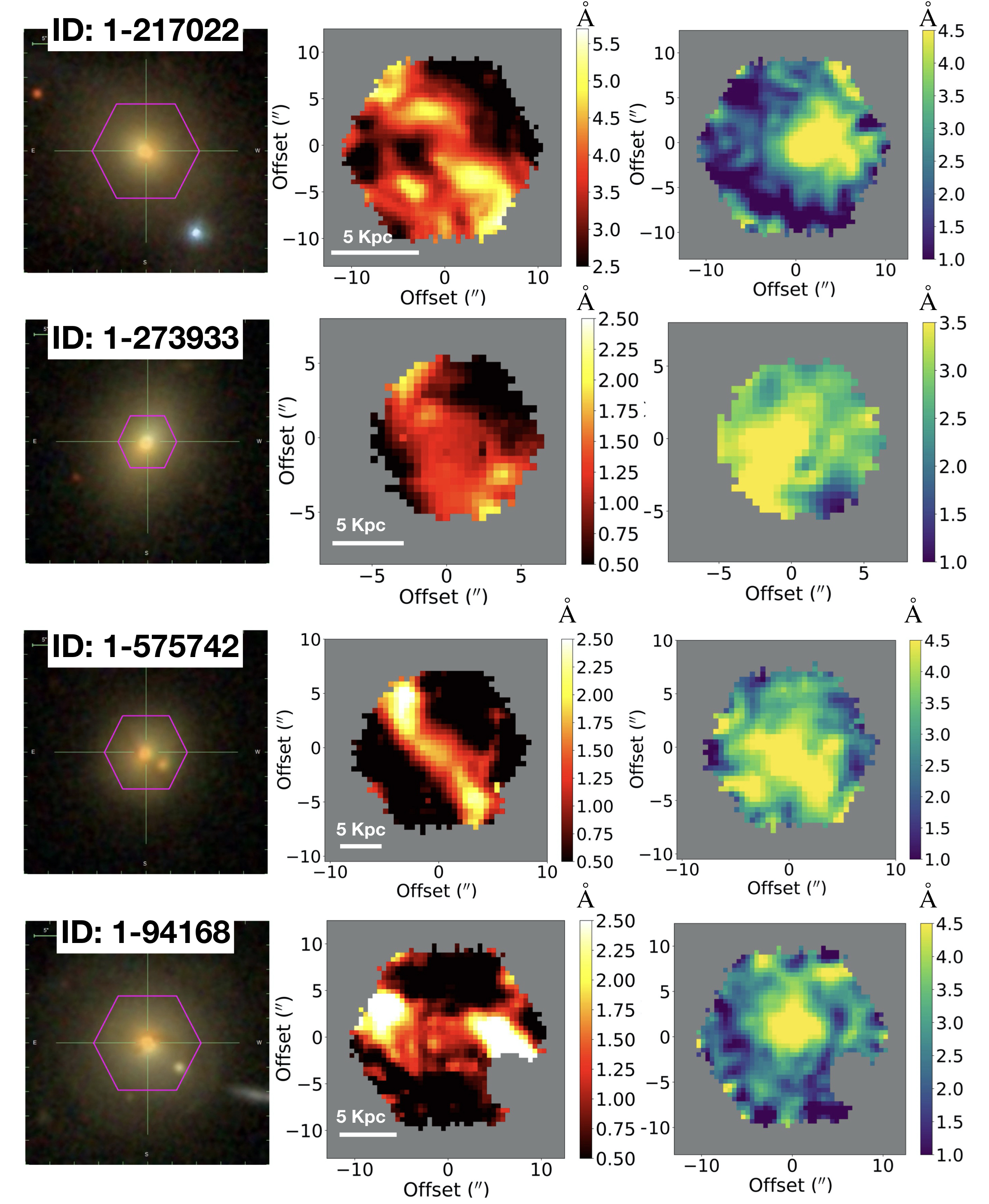}
\caption{  The spatially resolved emission and absorption line properties of 4 example red geysers (MaNGAID-1-217022, 1-273933, 1-575742 and 1-94168) from SDSS-IV MaNGA. The left panels show the optical images of the galaxies from SDSS with the MaNGA IFU overlaid on top in magenta. The on-sky diameter of the IFU fibers shown here range from 12$''$ to 22$''$ corresponding to physical size of roughly 10$-$20 Kpc.}  The middle panels show the H$\alpha$ Equivalent width (EW) maps (observed-frame units in \AA) showing the signature bi-symmetric pattern identifying the red geyser class, and the right panels show the EW map of NaD absorption (in \AA) which includes both stellar and cool ISM gas component. 
\label{fig:geysers_eg}
\end{figure*}

Red Geysers are early type galaxies (ETGs) lying in the red sequence ($\rm NUV-r>5$) which have little-to-no ongoing star formation activity. Fig.~\ref{fig:sfr} shows the star formation rates (SFR) of the red geyser sample (red circles) compared to all MaNGA galaxies (grey) from \cite{salim16} catalogue. This catalog derives SFR from ultra-violet/optical
spectral energy distribution (SED) fitting with additional constraints from mid-infrared data using GALEX-SDSS-WISE. The typical value of SFR for the red geysers is $\sim 0.01$ $\rm M_\odot~yr^{-1}$. The sample of red geysers used here is derived from MaNGA Product Launch 9 (MPL-9) and consists of 140 galaxies which accounts for $\approx6-8\%$ of the local quiescent galaxy population observed by MaNGA. The red geyser sample is visually selected based on their characteristic features, as described in detail in \cite{cheung16, roy18, roy20}. They are briefly outlined below:

\begin{itemize}[noitemsep]
\item Spheroidal galaxies with no disk component as observed by SDSS and low star formation rate.
\item Bi-symmetric feature in spatially resolved EW map of strong emission lines like H$\alpha$, [NII] and [OIII].
\item Rough alignment of the bi-symmetric feature with the ionized gas kinematic axis, but misalignment with stellar kinematic axis.
\item High spatially resolved gas velocity values, typically reaching a maximum of $\pm \rm ~300~km~s^{-1}$, which are greater than the stellar velocity values by atleast a factor of $\rm 4-5$.
\item High gas velocity dispersion values, reaching about $\sim 220 - 250 \  \rm km~s^{-1}$ in distinct parts of the galaxy.
\item Predominantly showing LINER or Seyfert type line ratios in the spatially resolved BPT diagrams. 
\end{itemize}

Four example red geysers are shown in Fig.~\ref{fig:geysers_eg}. The optical images (left panels) from SDSS show spheroidal morphologies typical of these galaxies. The middle panels show the characteristic bi-symmetric feature in the $\Ha$ EW map. The right panels show the NaD EW map obtained using MaNGA-DAP (see \S 10 in \cite{westfall19} for details). The DAP performs the spectral index measurements after first subtracting the
best-fit emission-line model from the observed spectrum. It does not perform any stellar continuum subtraction on the NaD feature, hence the observed EW shows the presence of both the stellar and the ISM component. However the NaD maps show highly anisotropic and clumpy spatial morphology, indicating that a major portion of the observed NaD absorption is likely contributed by ISM gas. We also find that the regions with high NaD EW are spatially offset from the bi-symmetric feature observed in $\Ha$ map suggesting the presence of multiphase gas components tracing different activities in these galaxies.   

\cite{roy18} cross-matched the red geyser sample with VLA-FIRST catalog to identify the radio detected red geysers. For the latest sample from MPL-9, we have an updated list of radio detected red geysers consisting of 30 galaxies.

We construct a control sample of spheroidal quiescent galaxies which are matched in global
properties, namely stellar mass, color, redshift and axis ratio, but do not show the resolved geyser-like features as described previously. For each red geyser, we match up to four quiescent galaxies (having $\rm NUV-r>5$) with the following criteria:

\begin{itemize}[noitemsep]
    \item log~M$_{\star , \rm red~geyser}$/M$_{\star ,\rm  control}<$ 0.2 dex 
    \item z$_{\rm red~geyser}$ - z$_{\rm control} <$ 0.01
    \item b/a$_{\rm red~geyser}$ - b/a$_{\rm control} < 0.1$
\end{itemize}

\noindent where M$_{\star}$ is the stellar mass, z is the spectroscopic redshift,
and b/a is the axis ratio from the NSA catalog. This technique results in 458 unique control galaxies with 65 of them detected in FIRST.  These constitute the radio-detected control sample. 

\begin{figure}[h!!!] 
\centering
\graphicspath{{./plots/}}
\includegraphics[width = 0.5\textwidth]{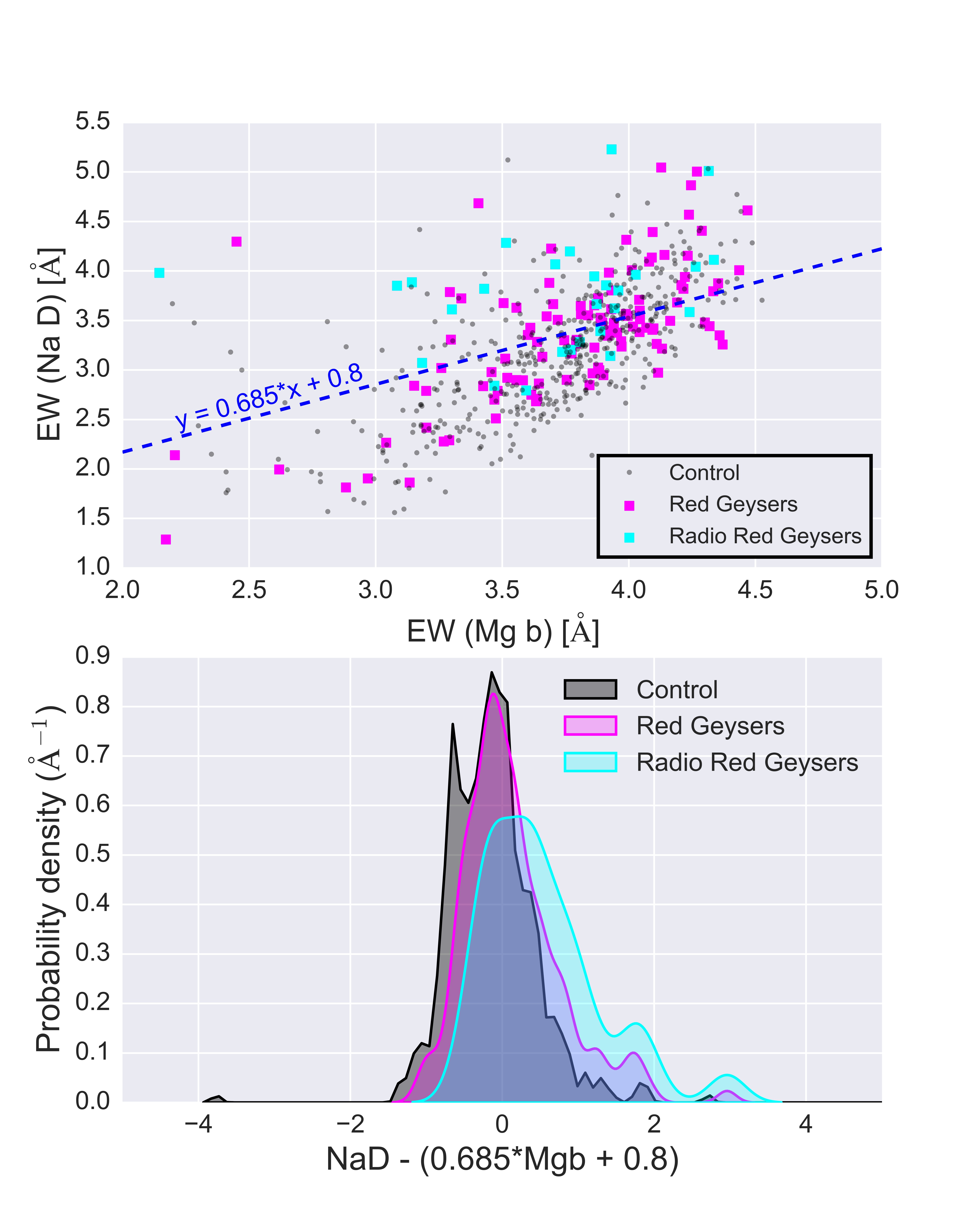}
\caption{ [Upper panel] Equivalent widths EW(NaD) vs. EW(Mg b) of the red geysers (magenta squares) and control sample (grey circles). The FIRST-detected radio red geysers are marked in cyan. The dashed blue line represents the empirical relation from stellar component only found in \cite{alatalo16} for star-forming galaxies. Any object above the relation is expected to have a dominant ISM contribution in NaD. [Lower panel] Probability density functions for the Na I D EW, measured with respect to the mean \cite{alatalo16} relation: EW(Na D) = 0.685*EW(Mg b)+0.8 (indicated by blue dashed line in the upper panel) for the different subsamples. The radio-detected red geysers (cyan) show a clear departure from the other two distributions, with a large fraction of objects showing excess NaD compared to their Mgb. 
\label{fig:nad_mgb}}
\end{figure}

\begin{figure}[h!!!] 
\centering
\graphicspath{{./plots/}}
\includegraphics[width = 0.5\textwidth]{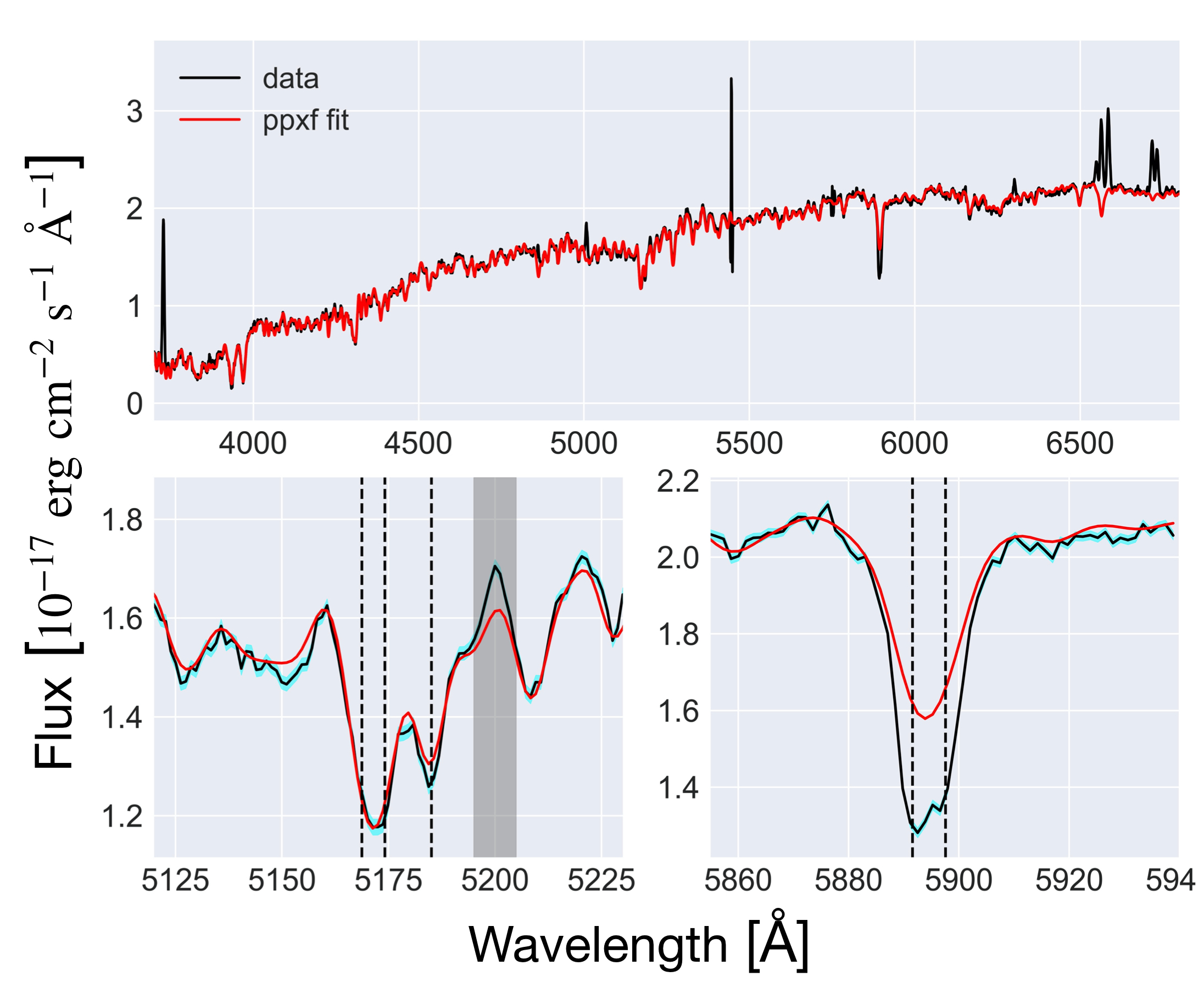}
\caption{ The top panel shows an example MaNGA spectrum in black. The best fit stellar continuum obtained from ppXF fit using MIUSCAT stellar population models, as described in \S\ref{stellar}, is overplotted in red. The result of the continuum fit around Magnesium triplet (Mg b) and Sodium doublet (NaD) absorption lines are shown in the left and right subplots in the bottom panel respectively. 1-$\sigma$ errors in the observed spectra are shown in the bottom panels in cyan. The gray shaded region in the lower left panel highlight the [NI] emission line at 5200 \AA. 
\label{fig:fullfit}}
\end{figure}

\section{Data Analysis} \label{sec:analysis}

\subsection{Stellar continuum fitting} \label{stellar}

The sodium doublet feature can arise from photospheric transitions in evolved cool stars, as well as from cool neutral gas in the ISM. Since our samples consist of red and quiescent galaxies with evolved stellar populations, the  presence of 
K-M type stars throughout the galaxy are likely to make up a major fraction of the spectral NaD profile \citep{jacoby84, tyson90, tyson91} with the residual absorption attributed to the ISM component. On the other hand, since the primary contribution in the Magnesium triplet MgI$\lambda$5167,5173,5184 \AA~(Mg b) absorption is from the stars, studies of star-forming galaxies have found a linear relation between the EW of Mg b and NaD from the stellar component alone  \citep{alatalo16}. Inspite of the observed scatter around the empirical relation in \cite{alatalo16}, any enhancement in NaD compared to Mgb beyond the relation likely indicates the ISM contribution. 

Fig.~\ref{fig:nad_mgb} (upper panel) compares the EW of NaD and Mgb of red geysers (in cyan and magenta squares, representing radio-detected and non-radio-detected galaxies from FIRST survey respectively) and the control sample (in grey circles), obtained by averaging the EW measured by the MaNGA DAP (that includes NaD absorption from both stellar and ISM component) over all spaxels within one effective radius. 
The figure shows that a large fraction of the galaxies in the radio-detected red geyser sample have a considerable excess in NaD lying above the \cite{alatalo16} empirical relation (blue dashed line). 

This is further confirmed by the lower panel in Fig.~\ref{fig:nad_mgb}, which shows the probability density function (PDF) of the deviation of NaD EW distribution with respect to the \cite{alatalo16} relation. We find that a large fraction of radio-detected red geysers show an excess amount of NaD, as indicated by the clear positive wing in the distribution, compared to the other two samples. 
Infact, 75\% of the radio-red geyser sample lie above the  relation, which compares to 51\% in non-radio detected red geysers and only 32\% in the control sample. This indicates an elevated abundance of cool gas within radio-detected red geysers and a possible correlation with the presence of radio-detected AGNs that we will return to below.

It is to be noted from Fig.~\ref{fig:nad_mgb} that there exist a notable amount of NaD deficit at low Mgb particularly for the control galaxies. This is likely caused by the in-filling of the resonant NaD emission line which has been previously detected in integrated, stacked as well as spatially resolved NaD spectrum \citep{alatalo16, chen10, rupke21}. The non-radio detected red geysers, on the other hand, show roughly an equal fraction of galaxies with NaD deficit (at low Mgb) and excess (at high Mgb) with a rather symmetric distribution overall. 

The next step is to accurately model out the stellar continuum contribution to the NaD absorption in order to extract the residual gaseous component. 
First, we mask spaxels with low or no spectral coverage, low signal-to-noise and having foreground stars, unreliable fits or noisy observations as flagged by MaNGA under the mask-bit names \texttt{NOCOV, LOWCOV, FORESTAR, UNRELIABLE} and \texttt{DONOTUSE}. Then, following Rubin et al. (in prep), we bin and stack the spectra in $\rm 2'' \times 2''$ spatial bins (or spaxels) to acquire a good continuum signal to noise (S/N$>$10 per dispersion element) per bin throughout the galaxy. 

We then fit the stellar continuum for each spatial bin with ppxf \citep[Penalized Pixel Fitting,][]{cappellari04, cappellari17} which fits a linear combination of simple stellar population (SSP) model templates. Here, we make use of the MIUSCAT SSP \citep{vazdekis12}, although a detailed comparison with other SSP models is presented later. 
Strong emission line regions are masked during the fit. We also mask the region around the NaD transition, as we assume that the doublet is a result of stellar + ISM contribution and hence should not be fit by the model. We also mask the red half of the He I emission line at 5875.67 \AA, which is close enough to the NaD line that it could affect the residual profile. An example spectrum and its continuum fit is shown in Fig.~\ref{fig:fullfit}. The stellar continuum model shows a satisfactory fit, with minimal residuals around Mg b and an excess around NaD, which hints at the presence of ISM gas.

To remove the stellar continuum and isolate the ISM component, the spectrum in each bin is divided by the best corresponding continuum fit. We additionally fit a first-order polynomial to the continuum-normalized spectrum in the wavelength range immediately (20 \AA) blueward and redward of the NaD profile and divide the residual by the polynomial fit. This is to account for any systematic errors in continuum fitting that might give rise to artificial residuals. 

\begin{figure}[h!!!] 
\centering
\graphicspath{{./plots/}}
\includegraphics[width = 0.49\textwidth]{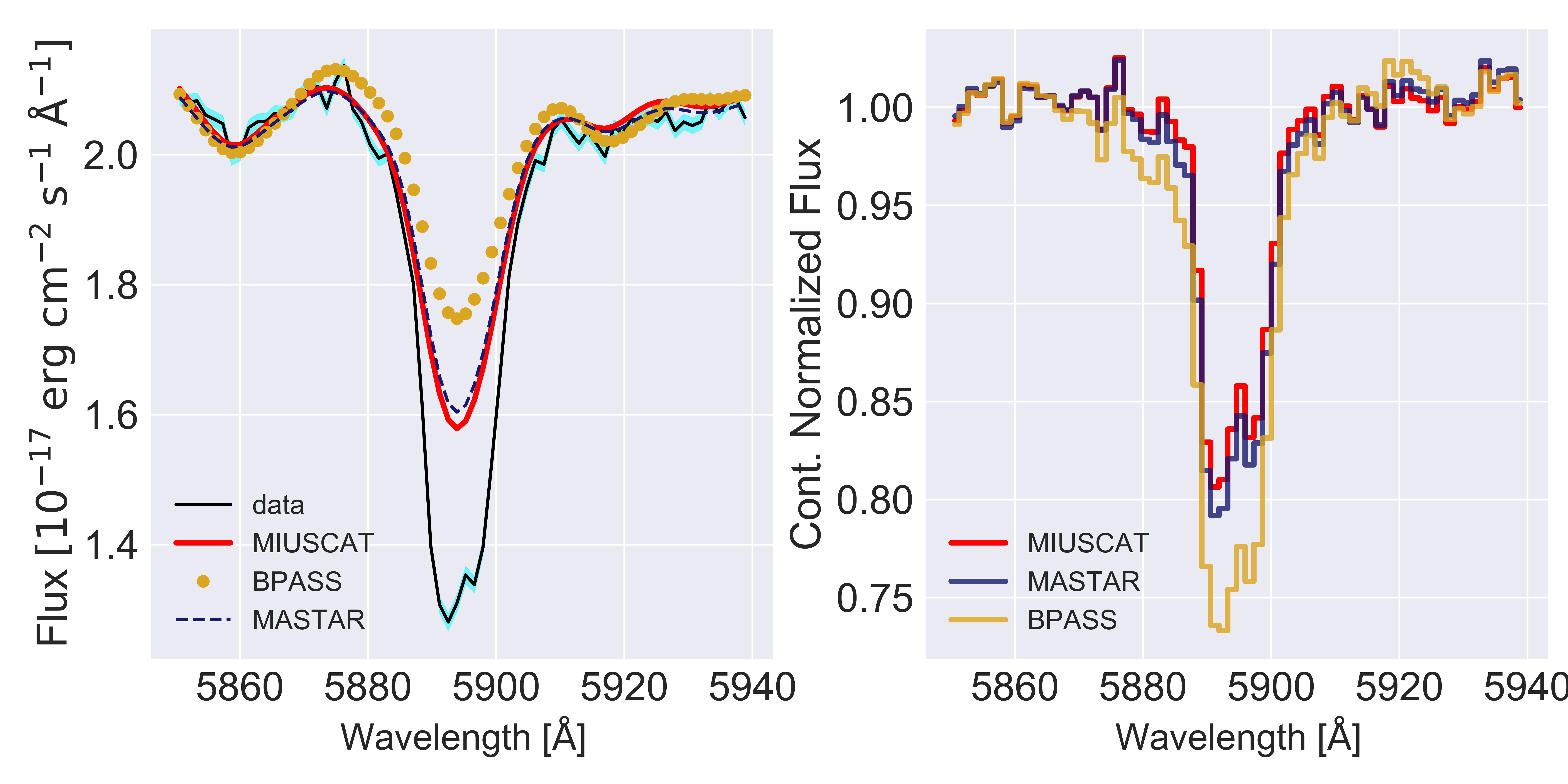}
\caption{ Left panel shows the modeled stellar continuum obtained by using three different stellar population models (MIUSCAT, BPASS and MASTAR) in different colors and line styles, overplotted on the observed NaD absorption spectra in black. 1-$\sigma$ error of the observed NaD spectrum is shown in cyan. Right panel shows the corresponding continuum normalized residual NaD indicating contribution from the ISM.
\label{fig:contfit}}
\end{figure}

We now examine the robustness of our stellar continuum modeling to the assumed stellar library.  We repeat the above continuum fitting process for 10 galaxies in each of red geyser and control sample across every spatial bin using two additional libraries, namely MaStar \citep{mastar}, an empirical library constructed using the MaNGA instrumentation, and BPASS \citep{bpass}, a theoretical template library.  The BPASS SSP models are drawn from the v2.2.1 release that adopted the default BPASS initial mass function\footnote{These are freely available at https://bpass.auckland.ac.nz/9.html}. SSPs over a range of log age (in Gyr) from $\rm -1.2$ to $1.1$, with stellar metallicity mass fractions Z=0.008, 0.02, and 0.03, are included. The BPASS fitting will be described in full in Parker et al. (in prep).

An example of the variation of the stellar contribution owing to the different stellar population models is shown in the left panel of Fig~\ref{fig:contfit}. The different colors indicate the SSPs used to construct the stellar model. The right panel shows the continuum normalized NaD spectra using those same SSPs. We find that the fractional difference in NaD EW of the continuum-normalized spectra, computed using the MASTAR and the MIUSCAT models, range between 5-15 \% in the ten red geyser galaxies studied. However, the BPASS model can change the estimated EW by about 25-30\%, compared to the other two models. The BPASS is the result of combining theoretical model spectra with as few empirical inputs as possible. Hence this differs largely from the two other stellar models (MIUSCAT and MASTAR) which are based on empirical libraries. Preliminary work has revealed that the BPASS continuum model may be under-predicting the amount of NaD coming from stellar atmospheres (Parker et al. in prep). On the other hand, the stellar models from MIUSCAT and the MASTAR models, drawing on empirical libraries, likely contain some interstellar NaI absorption from the Milky Way, overpredicting NaD absorption from stars. Thus, the three models are different and are shown to explore a broad range of parameter space in stellar continuum modeling and their effect in our analyses. 

Although the equivalent width value of the continuum normalized NaD spectra changes owing to the different stellar models, the velocity difference is negligible ($\rm <~5~km~s^{-1})$, as shown by the example in Fig~\ref{fig:contfit}. Among the three libraries we tested, we use the MIUSCAT SSPs in what follows because this particular set of models yields the most conservative (lowest) value of inferred NaD ISM contribution.


\subsection{Equivalent Width Calculation} \label{ew}

Once the spectrum is continuum-normalized, the strength of the NaD absorption arising from the ISM is quantified by the equivalent width (EW). EW is measured as follows:

\begin{equation}
    \rm W= h_{\lambda} \sum_{j = 1}^{M} \left[1 - \frac{F_j}{F_c}\right]
\end{equation}

\noindent where h$_{\rm\lambda}$ denotes the dispersion in wavelength in \AA/pixel, M is the number of individual pixels that together constitute the desired wavelength interval which includes 20~\AA~immediately blueward and redward surrounding and including the NaD absorption feature, and F$_{\rm j}/F_{\rm c}$ is the continuum normalized flux for each of those pixels.

\subsection{NaD line fitting using Bayesian inference} \label{fitting}


After continuum-normalizing the spectra as discussed in \S\ref{stellar}, we want to determine the kinematics of the neutral gas present in our galaxy samples. 
We perform a ``detection'' test for every spaxel to determine the significance of the ISM component in the NaD spectra with the following criteria:
\begin{enumerate}[noitemsep]
    \item The depth of the absorption trough in the residual spectra must be at least $\sim \rm 10\%$ of the continuum level.
    \item The equivalent width measured from the residual ISM component must be $>0.3$~\AA.
    \item The S/N of each spectrum $>10$ per dispersion element.
\end{enumerate}
  The primary reason behind taking such restrictive criteria is to prevent biases from low signal-to-noise (S/N) spectra with extremely small inferred residuals which would provide inaccurate kinematic estimates from erroneous model fits. As discussed later, input spectrum with $\rm S/N < 10$ gives an error of up to $\rm \sim 20~km~s^{-1}$ in measured velocities which can be marginally close to the observed velocity amplitudes we wish to investigate. Requiring spectra with $\rm S/N > 10$ ensures that the velocity errors remain within acceptable limits. 
  
  The NaD spectral feature in the spaxels which satisfy all the requirements are then modelled with an analytical function \citep{rupke05}, which takes the following form:

\begin{equation}
    I(\rm \lambda) = 1 - C_{f} + C_{f} \times e^{-\tau_{\rm B}(\lambda) - \tau_{\rm R}(\lambda)}
\end{equation}

\noindent where C$_{f}$ is the velocity-independent covering factor, and $\rm \tau_{B}(\lambda)$ and $\rm\tau_{R}(\lambda)$ are the optical depths of the Na I $\rm\lambda$ 5891~\AA~and Na I $\rm\lambda$ 5897~\AA~(vacuum wavelength) lines, respectively. The optical depth of the line, $\rm\tau(\lambda)$, can be expressed as:

\begin{equation}
    \tau(\rm \lambda) = \tau_{0} \times e^{-(\lambda - \lambda_{0} + \Delta\lambda_{\rm offset})^{2}/((\lambda_{0} + \Delta\lambda_{\rm offset})b_{\rm D}/c)^{2}}
\end{equation}

\noindent where $\tau_{0}$ and $\lambda_{0}$ are the central optical depth and central wavelength of each line component, respectively, b$_{D}$ is the Doppler line width, and c is the speed of light. The wavelength offset is converted from a velocity offset, given by $\Delta\lambda_{offset}$ = $\Delta v \lambda_{0}/c$. For NaD, $\tau_{0,B}/\tau_{0,R}$ = 2. The optical depth parameter can be derived from the column density of sodium described as :

\begin{equation}
    N(\rm Na I) = \frac{\tau_{0}b}{1.497 \times 10^{-15} \lambda_{0}f}
\end{equation}

where $\lambda_{0}$ and f are the rest frame wavelength (vacuum) and oscillator strength, respectively. Throughout this study we assume $\lambda_{0}$ = 5897.55Å and f = 0.318 \citep{morton91}.

\begin{figure}[h!!!] 
\centering
\graphicspath{{./plots/}}
\includegraphics[width = 0.5\textwidth]{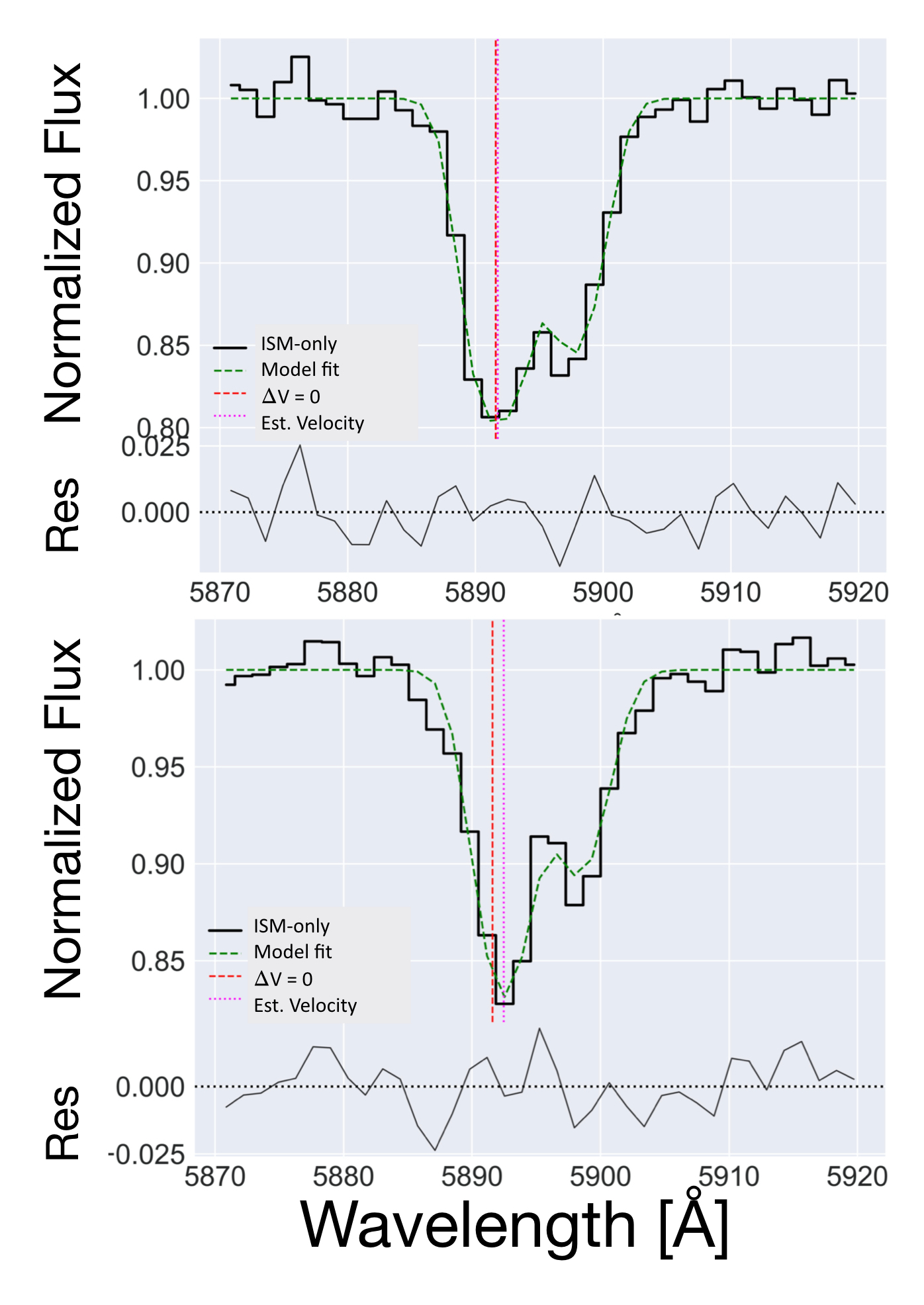}
\caption{  The two panels show the modeling of continuum-normalized NaD spectra using \cite{rupke05} absorption model for two different spaxels of a red geyser (MaNGA ID: 1-217022). The observed spectra are shown in black, the best fit models in green, and the wavelengths corresponding to the estimated velocities in magenta. The bottom subpanels below the main plots show the residuals of the fits.
\label{fig:modelfit}}
\end{figure}

The absorption feature is fitted here using a single kinematic component. Fitting the neutral gas absorption with one component is certainly an over-simplification in a scenario where contributions from more than one gas cloud along the line of sight are embedded within a single line profile. However, this approach allows us to characterize the global kinematics obtained from the data. The absorption model is convolved with MaNGA's instrumental resolution ($\rm \sim 65~km~s^{-1}$) before performing the fit.

In order to explore possible degeneracies, obtain unbiased fits, and estimate the errors, we wrap our fitting procedure in a Bayesian inference approach using the Dynamical Nested Sampling algorithm \citep{higson19} as implemented in the Python package, \textit{dynesty} \citep{speagle19}. Nested Sampling  \citep{skilling06, skilling04} estimates both the Bayesian evidence and posterior distributions in an iterative fashion until the convergence criteria is met. The method has the flexibility to sample from complex, multi-modal distributions using adaptive sampling to maximise the accuracy and efficiency of the fitting process. The greater flexibility and the ability to compute Bayesian evidence (Z) on top of posteriors are advantages compared to traditional Marcov Chain Monte Carlo (MCMC) methods. The default stopping criterion for \textit{dynesty} is used in our analyses with no limitation on the maximum number of iterations. We have set reasonable priors in our MCMC fitting routine: the line width is constrained within $\rm \pm 100~km~s^{-1}$ around the stellar line width obtained from the continuum model and the limits on the allowed velocity are set at $\rm \pm 500~km~s^{-1}$. The covering fraction takes values within 0 and 1. No constraints are placed on the column density parameter. 
Fig.~\ref{fig:modelfit} shows two examples of model fits applied to two different spatial bins of a red geyser (MaNGA ID: 1-217022). 
The residuals vary from $2-5\%$, indicating an acceptable model fit.

For each input spectrum, our procedure delivers estimates of the column density, velocity, line width, and the covering fraction. The best fit velocity with respect to the systemic velocity is assigned to the kinematics of the neutral ISM at that location.  

\begin{figure}[h!!!] 
\centering
\graphicspath{{./plots/}}
\includegraphics[width = 0.5\textwidth]{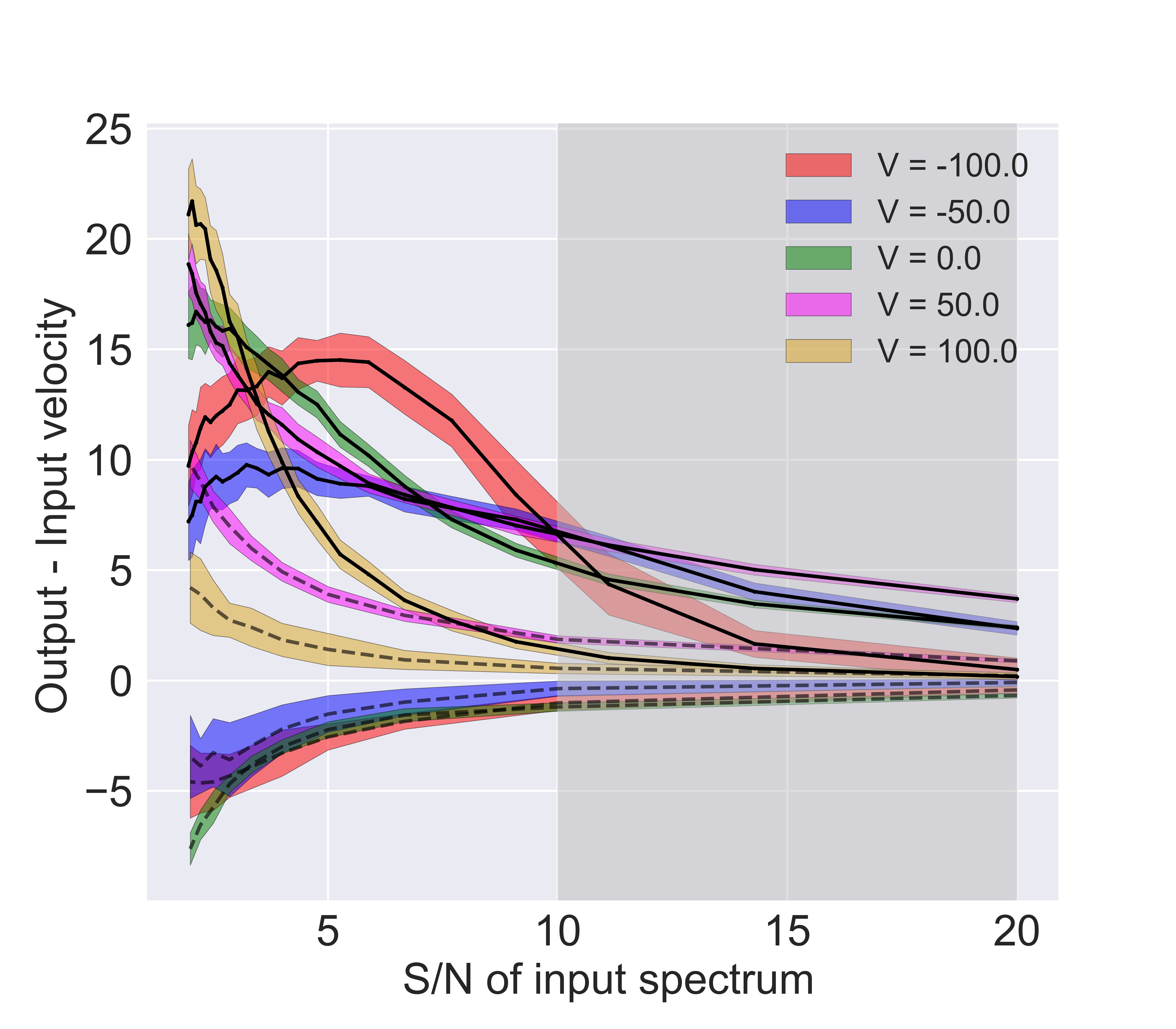}
\caption{Difference between the input velocity (V) of the ISM component used to simulate the mock NaD absorption spectra and the retrieved velocity obtained with our velocity estimation analysis vs. the signal-to-noise ratio used in the mock simulation. Each color represents 'true' ISM velocities used for simulating the data. Colored shaded regions depict 1$\sigma$ error regions on the difference. The solid lines are for realistic sodium doublet absorption profiles while dashed lines are for doublet profiles which are mirror images of the former. Grey shaded region indicates the S/N in our observed spectrum. This shows the bias in our velocity retrieval process.
\label{fig:sim}}
\end{figure}

As a key aspect of this paper is a characterization of the global NaD kinematics in our red geyser sample, it is important to test the robustness of our velocity measurements for systematic biases.  Such biases might be expected: we are fitting an asymmetric absorption doublet model to the spectral residual obtained after subtracting a broadened and possibly offset version (from the stellar continuum) of a similar doublet model.  

We begin by generating a synthetic stellar NaD absorption spectra assuming a Gaussian stellar velocity profile of amplitude and width as observed from the stellar continuum model of our example galaxy with MaNGAID$-$1-217022. 
The mean of the Gaussian is set to the systemic velocity of the galaxy. We simulate the ISM component using the doublet absorption model of \citet{rupke05} with average values set for each parameter, namely, column density$= \rm 2.9 \times 10^{13}~cm^{-2}$,  line width ($\sigma$) $= \rm 150~km~s^{-1}$ and covering fraction $= \rm 0.2$.  These parameter values were obtained by fitting the observed NaD-ISM spectra for the same galaxy with the same \citet{rupke05} model.  Mock spectra are created for five assumed input velocity offsets (-100 $\rm km~s^{-1}$, -50 $\rm km~s^{-1}$, 0 $\rm km~s^{-1}$, 50 $\rm km~s^{-1}$ \& 100 $\rm km~s^{-1}$) relative to the systemic velocity.  We convolve the resultant profile with the instrumental resolution of MaNGA and then add normally-distributed noise to the simulated spectra. We perform the above process for different signal-to-noise values ranging from 1 to 20 at each input velocity. We fit the resultant spectra and retrieve the model parameters according to the technique described above and in \S\ref{stellar}. The error estimates on the retrieved velocities are obtained by bootstrapping the simulated `data'. 

The difference in the retrieved versus input velocities for different values of signal-to-noise ratios are shown in the Fig.~\ref{fig:sim} via solid lines. Each colored line represents a different initial velocity offset value. The plot reveals systematic biases that result from our model fitting and velocity retrieval process.  We see that a modest positive (redshifted) bias exist, but for at $\rm S/N>$10 per dispersion element (grey shaded region), similar to the noise level in the NaD spectra used in our analyses, the bias in all cases is less than $\sim\rm 5~km~s^{-1}$.  As we show, this systematic error is roughly an order of magnitude smaller than the average velocity offsets we detect in our red geyser samples.

Such a bias probably arises due to the combination of the inherent shape of the subtracted stellar component as well as the interstellar component of the NaD doublet feature. To check that, we produce similar mock NaD spectra as before, but this time we reverse the shape of the doublet feature, while keeping the stellar absorption model same. We perform a similar velocity retrieval process and the results are shown in dashed lines. We see that the bias has now switched to the negative side for input velocities $< 0$ and show a much smaller positive bias compared to the previous exercise for positive velocities. 
Varying one or a combination of other parameters like column density, covering fraction or line width has no impact on the amplitude of the bias.

\subsection{Stacking optical spectra and interpretation of stacked kinematics}

In what follows, we will be interested in the average behavior of the NaD ISM component across various subsamples as a check on our model fits of spatially-binned spectra in individual galaxies.  To compute spatially-integrated spectra per galaxy, we shift the spectra from each spaxel to the rest frame of the galaxy and continuum normalize them before coaddition to remove the stellar contribution by method discussed in \S\ref{stellar}. After generating a pure-ISM spectra from each spaxel, we construct a circular radius of 8$''$ in each galaxy (which is roughly the angular scale corresponding to the average effective radii of our targets) and include only those spaxels within the radius which satisfy the detection criteria. This is done to remove low signal-to-noise spaxels as well as spaxels near the extreme edge of the IFU. We co-add the spectra and perform weighted averages where the weights are the S/N of each spectrum used in the stacking. 
The spatially integrated velocity thus obtained represents the average kinematics from the particular galaxy. Once we obtain an integrated spectrum for each galaxy, we also compute a stacked spectrum across each galaxy sample.


\section{Results} \label{results}

\begin{figure}[h!!!] 
\centering
\graphicspath{{./plots/}}
\includegraphics[width = 0.45\textwidth]{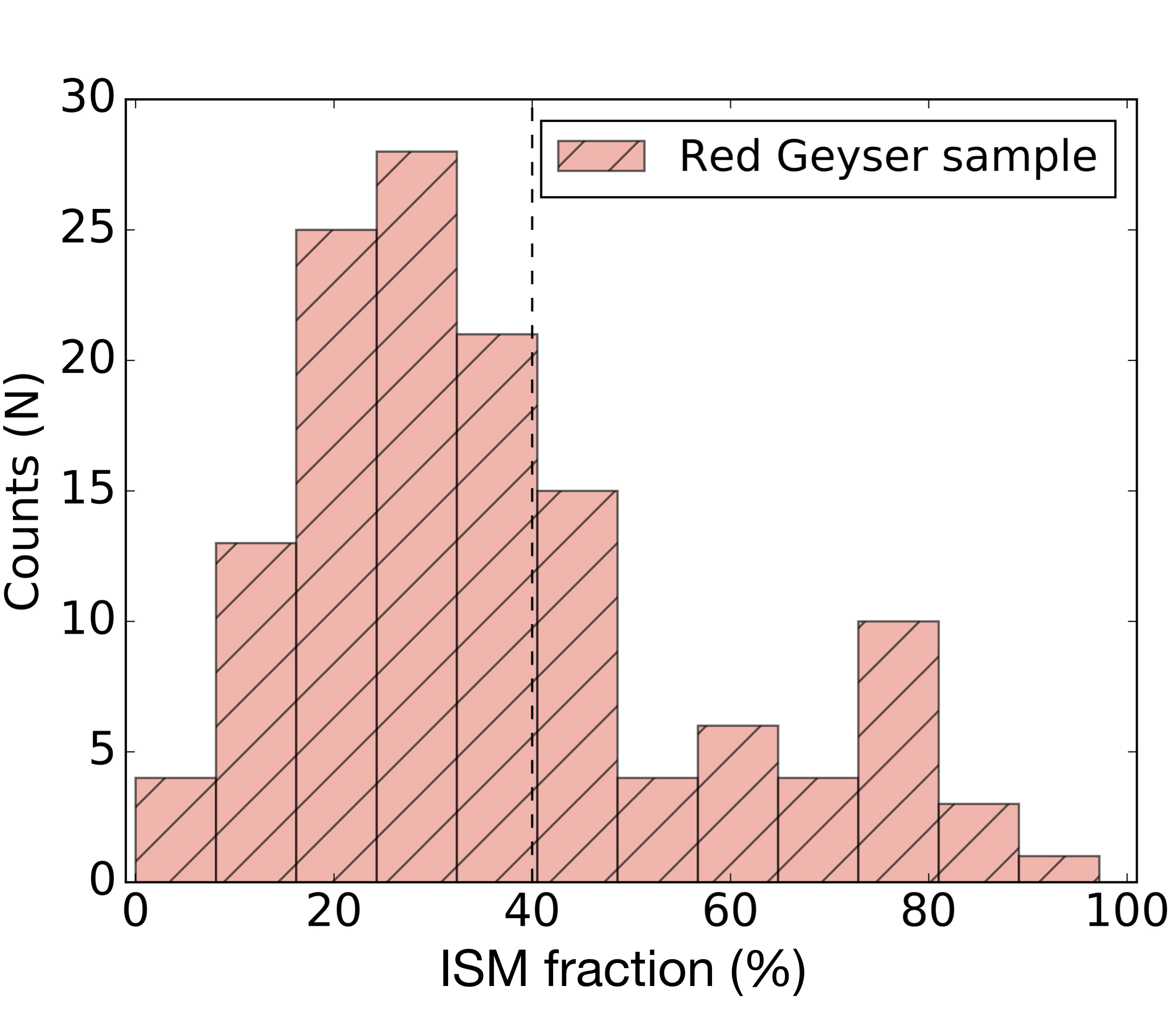}
\caption{  The distribution of mean percentage of ISM contribution in the NaD doublet absorption profile as observed in the red geyser galaxy sample. Dashed line separates red geyser galaxies with $\rm ISM \ fraction > 40\%$, which is greater than twice the typical fraction observed in control galaxies. 
\label{fig:fraction}}
\end{figure}

\begin{figure}[h!!!] 
\centering
\graphicspath{{./plots/}}
\includegraphics[width = 0.5\textwidth]{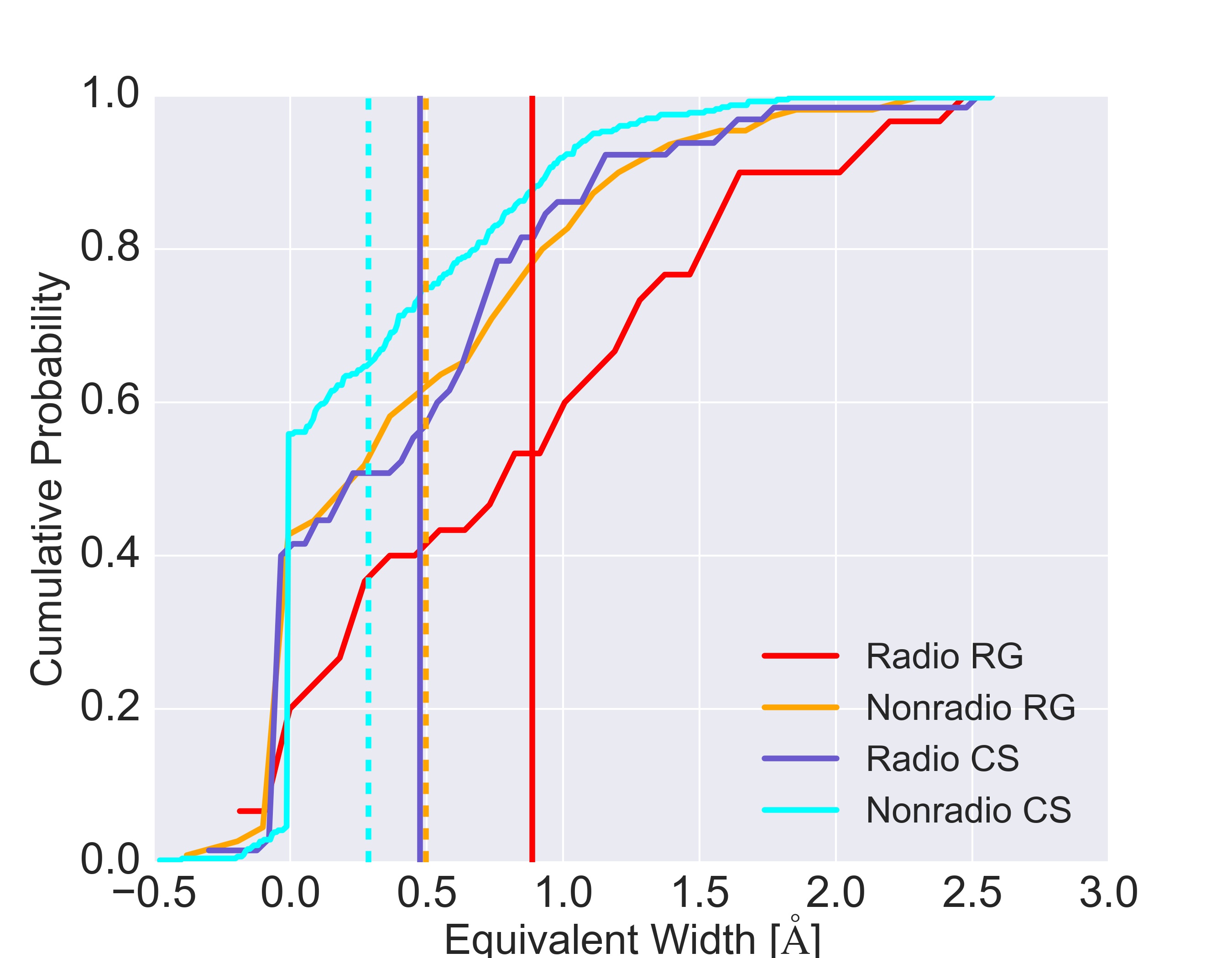}
\caption{ Cumulative distribution function showing the comparison of the distribution of mean NaD-ISM EW (computed by averaging over a circular radius of 8$''$) between four different samples} - radio detected red geysers (red), non radio detected red gesyer (orange), radio detected control (blue) and non radio detected control (cyan) galaxies. The mean of each distribution is shown by same-colored solid (radio-detected galaxies) and dashed (non radio detected samples) lines. The
radio detected red geyser shows the highest mean EW of $= 0.9$ \AA~and quite different distribution than control samples. 
\label{fig:meanew}
\end{figure}

\begin{figure}[h!!!] 
\centering
\graphicspath{{./plots/}}
\includegraphics[width = 0.49\textwidth]{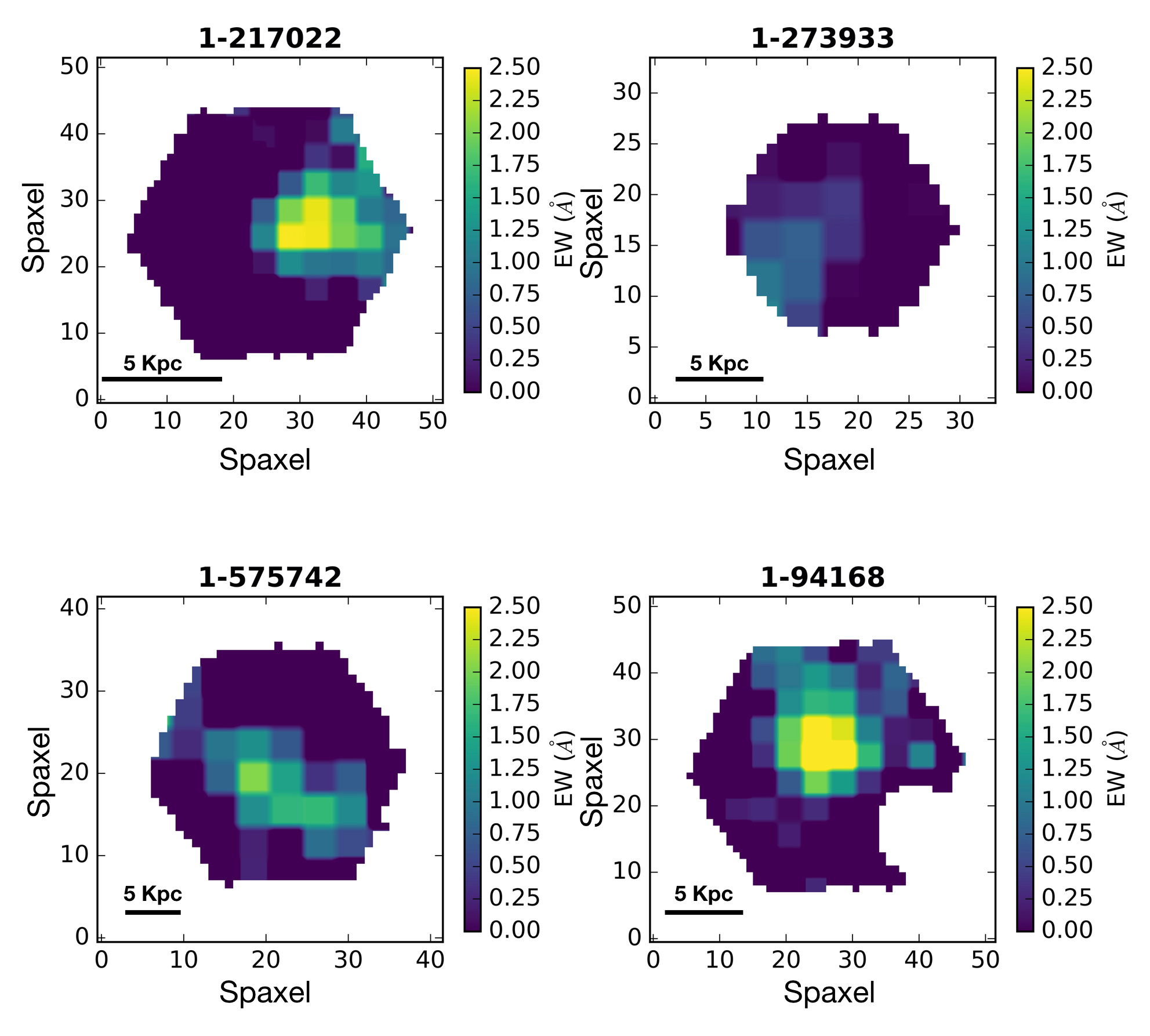}
\caption{  Spatially resolved NaD EW maps for four red geysers (MaNGA ID: 1-217022, 1-273933, 1-575742 and 1-94168) derived from the spaxel-wise continuum-normalized NaD spectra for each galaxy. Since the stellar continuum has been factored out, the displayed EW comes entirely from the ISM component. Out of the four red geysers shown here, three are radio detected (MaNGA ID: 1-217022, 1-273933, 1-94168). 
\label{fig:ismmaps}}
\end{figure}


With the ISM component of the NaD absorption spectra extracted, we derive estimates of the equivalent width, column density, and kinematics to search for potential differences between red geysers and normal galaxies.   
Our four samples of interest include 30 radio-detected red geysers (radio-RG), 110 non-radio-detected red geysers (nonradio-RG), 65 radio-detected control sample galaxies (radio-CS) and 393 non-radio-detected control galaxies (nonradio-CS).  We will compare these samples to investigate the frequency and strength of ISM NaD absorption in red geysers, both with and without radio AGN.  Finally we will investigate the NaD kinematics.


\subsection{Fractional contribution of Interstellar NaD absorption in red geysers} \label{fraction}

We begin by evaluating the fractional contribution of the ISM component to the total NaD absorption in our sample of 140 red geysers.  After fitting the stellar continuum according to the method described in \S \ref{stellar}
we calculate the mean EW of the continuum normalized NaD-ISM spectra for each red geyser galaxy by averaging over those spaxels that satisfy the 'detection test' described in \S\ref{fitting}. We then compute the ratio of the NaD EW from the ISM component to the total EW derived from the original spectra. The distribution of this ratio for the entire red geyser sample is presented in Fig.~\ref{fig:fraction}. 

Since passive early type galaxies are dominated by evolved stellar population which is the major contributor to the NaD absorption feature, the fraction of ISM component in the NaD EW is generally observed to be low. The fraction is $<\rm 20$\% in majority (80\%) of the control galaxies. 
On the other hand, we find that in 50 out of 140 red geysers (i.e., $35\%$ of the sample), the fractional ISM contribution is $> 40$\%, i.e. greater than twice the typical fraction observed in the control galaxies. 77 red geysers (i.e., $55$\% of the sample)  have ISM fraction $>$ 30$\%$. If we consider the spatially resolved map for each galaxy, roughly $\rm  56\%$ of the red geyser sample show an appreciable presence of ISM gas(i.e., spaxels satisfies the detection criteria in \S \ref{fitting}) over a projected area of $ \geq \rm 5\times5~kpc^2$. The fraction increases to $78\%$ if we consider only the radio detected red geyser sample. This compares to a mere $ 25\%$ for the entire control sample and  $ 41\%$ for the radio detected control galaxies.

\subsection{Comparison of mean EW in radio detected red geysers vs. control sample} \label{meanew}

Having shown that detections of substantial ISM components are 2--3 times more common among red geysers, we now investigate whether this implies that the total amount of NaD-associated gas, as seen in absorption, is also greater.  After removing the stellar contribution, we calculate the column density weighted mean NaD EW across all galaxies in the FIRST-detected radio-RG and nonradio-RG samples and compare them to radio-CS and nonradio-CS galaxies.

The comparison is shown as cumulative distribution functions in Fig.~\ref{fig:meanew}. The distribution of the NaD EW from the radio-RG galaxies appears to be different than both the radio detected and non-detected control samples. The nonradio RG, however, show a very similar distribution of NaD EW with the radio-CS sample. We perform a Kolmogorov-Smirnov (KS) test to measure the statistical significance of the differences of radio-RG sample with radio-CS and nonradio-CS samples and find that the null hypothesis that the said distributions are similar are rejected at a level  $<1\%$ with p values = 0.008 and $\rm 7\times 10^{-5}$ respectively. Additionally, the radio-red geysers have the highest mean with $\rm EW = 0.9^{+0.31}_{-0.30}$ \AA, greater by about 0.45~\AA~ than the radio-CS sample which has a mean $\rm EW = 0.46^{+0.29}_{-0.28}$ \AA~and more than the nonradio-CS ($\rm EW = 0.27^{+0.41}_{-0.26}$~\AA). The nonradio-RG also has a high NaD EW than both the control samples with a mean $\rm EW = 0.51^{+0.35}_{-0.33}$~\AA. This indicates a greater level of NaD-absorbing material in the ISM of the red geysers generally, but particularly so for the radio detected sample. 

Fig.~\ref{fig:ismmaps} shows the spatially-resolved NaD EW maps from the ISM components in four representative red geysers (MaNGA-ID: 1-217022, 1-273933, 1-575742, 1-94168) with ISM contributions $\rm \frac{EW_{ISM}}{EW_{tot}} > 30-40\%$. 
While all four of them exhibit an ISM NaD enhancement, EW in two of them even exceed 2.5~\AA~ in certain regions of these galaxies. 



\subsection{Spatial extent and morphology of the NaD feature} \label{sec:onsky_area}

\begin{figure}[h!!!] 
\centering
\graphicspath{{./plots/}}
\includegraphics[width = 0.5\textwidth]{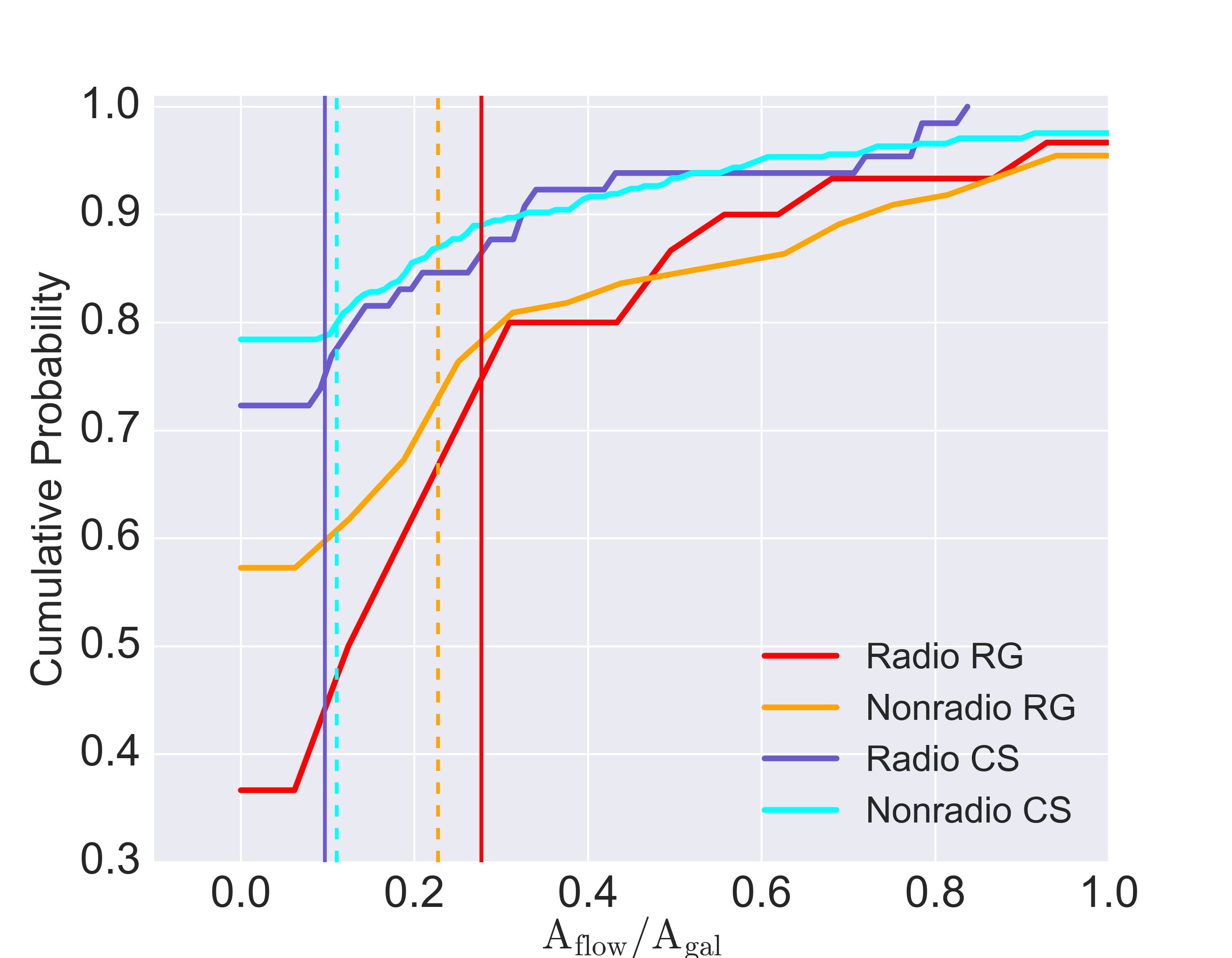}
\caption{ Cumulative distributions showing the fractional area of the galaxy comprising the cooler neutral ISM gas via NaD absorption for radio detected and not-detected red geysers (red and orange) and control samples (blue and cyan) galaxies respectively. Radio detected red geyser covers, on average, the highest fractional area of NaD absorption (red solid line), almost three times higher than that of the mean of radio detected control sample (blue solid line). 
\label{fig:areaonsky}}
\end{figure}

Fig.~\ref{fig:ismmaps} illustrates several common features in the spatial morphology of the ISM NaD distributions observed within red geysers.  These components typically have irregular shapes and are not smooth and symmetric, unlike what is typically expected from stellar distribution \citep[similar to ][]{cazzoli14}.
These asymmetric spatial distributions reinforce the ISM origin of this enhanced NaD absorption.

We have shown that ISM-associated NaD absorption is more common and globally stronger in red geysers, but we can also characterize the typical density distributions of this absorption to distinguish between dense, but relatively concentrated components versus those that are more extended and diffuse.  To quantify the spatial morphology and extent of the NaD absorption in each galaxy, we convert the number of spaxels comprising the NaD enhancements in each galaxy to the on-sky  projected area using the following conversions:

\begin{equation} \label{anad}
    \rm A_{flow} = A_{NaD} = N \times (S_{p})^{2}
\end{equation}

\begin{equation}
    \rm A_{gal} = \pi~R_{e}^{2}~(b/a)
\end{equation}

where N is the number of spaxels, $S_{\rm p}$ is the pixel scale which is 0.5$''$ for MaNGA, $R_{\rm e}$ indicates the effective radius of the galaxy and b/a gives the axis ratio.

Hence the quantity $\rm A_{flow}/A_{gal}$ gives the fractional area covered by the NaD ISM absorption compared to the on-sky projected area of the entire galaxy. Fig.~\ref{fig:areaonsky} shows the cumulative distributions of this fraction for radio-RG, nonradio-RG, radio-CS and nonradio-CS galaxies. We see that when ISM NaD absorption above our threshold is detected, it covers a wider fractional area ($0.28^{+0.09}_{-0.07}$) within the average radio-detected red geyser sample compared to its areal fractions ($\rm0.09^{+0.15}_{-0.09}$ and $\rm 0.1^{+0.1}_{-0.09}$) in the radio-detected and non-detected control samples respectively. This is consistent with \cite{rupke21} where the percentage of galaxy disk area showing inflows/outflows via NaD signature is found to be within 10-25\% for a sample of Seyfert galaxies.  KS test reveals that the distribution of fractional area from radio-RG sample is similar to that of nonradio RG sample (p value = 0.161) but differ significantly from the radio-CS (p $ = \rm 5\times 10^{-4}$) and nonradio CS (p $= \rm 1\times10^{-5}$) galaxies. Here, p value $< 0.01$ indicates the null hypothesis of the distributions being similar are rejected at a level $<$1\%. 
If we define a column density threshold of log N(Na I)$\rm/cm^{-2} > 12.3 $, which corresponds to the average NaD column density across the entire red geyser and control sample, we find that $\rm 63\%$ of radio-RG show NaD absorption which covers atleast $> 10\%$ of the on-sky projected area in the galaxy. This compares to $\rm 42\% $ in non radio-RG sample, $\rm 27\%~ and~21\%$ in radio-CS and nonradio-CS galaxies respectively.  Thus, the spatial extent of the cool gas present in the FIRST-detected radio enhanced red geysers far exceeds those from the other samples of galaxies.

\subsection{Spatially resolved NaD kinematics} \label{resolved_kin}

\begin{figure}[h!!!] 
\centering
\graphicspath{{./plots/}}
\includegraphics[width = 0.5\textwidth]{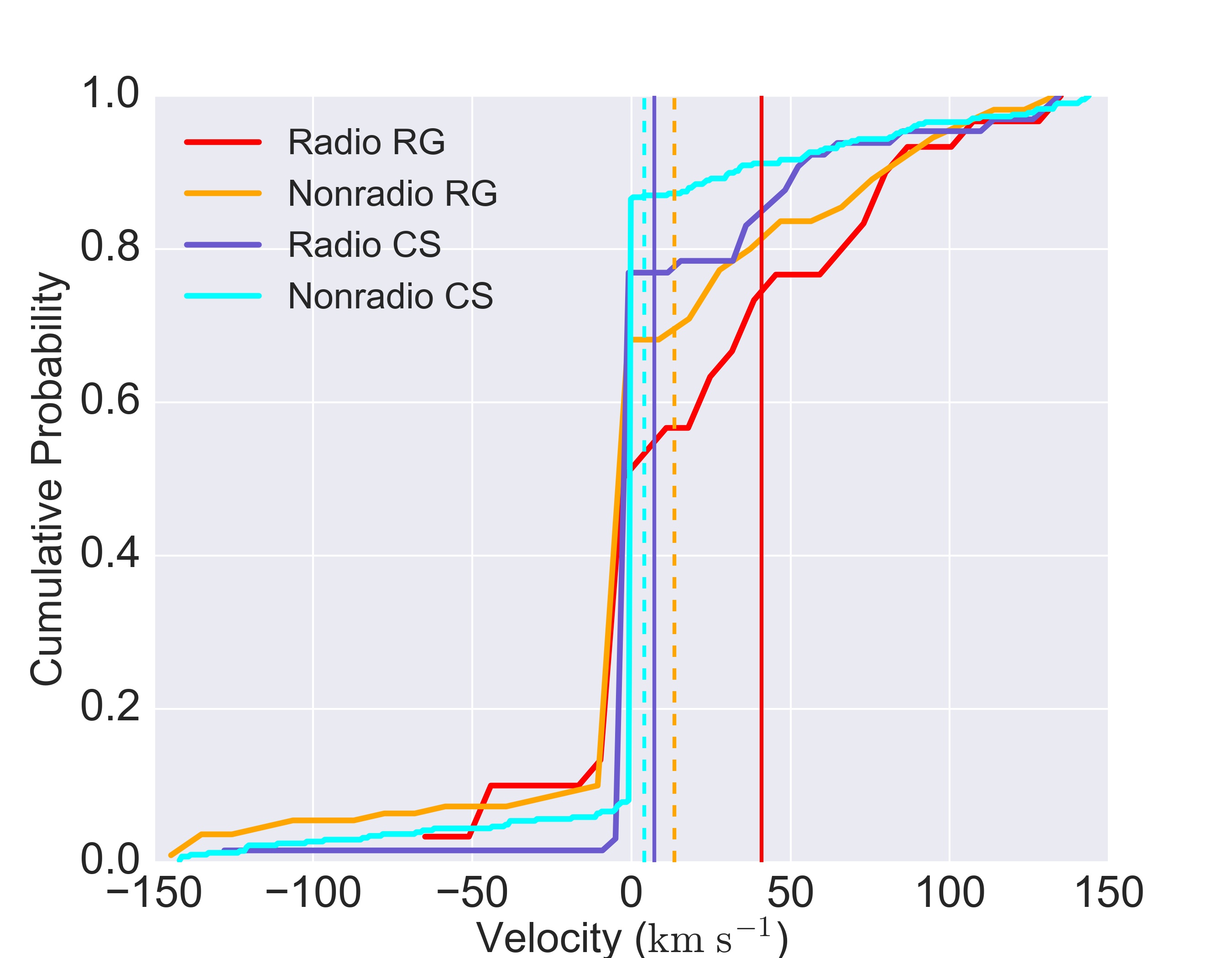}
\caption{ The cumulative distribution of mean velocities obtained from spatially resolved NaD kinematics for red geyser and control samples, split by radio detection. The color scheme is the same as Fig.~\ref{fig:meanew}. The mean velocities are calculated by computing the average of spatially resolved velocities over a circular radius of 8$''$. Both the radio detected and non-detected red geyser samples shows a clear excess  in redshift or a positive velocity compared to control samples.
\label{fig:intvel}}
\end{figure}

\begin{figure}[h!!!] 
\centering
\graphicspath{{./plots/}}
\includegraphics[width = 0.5\textwidth]{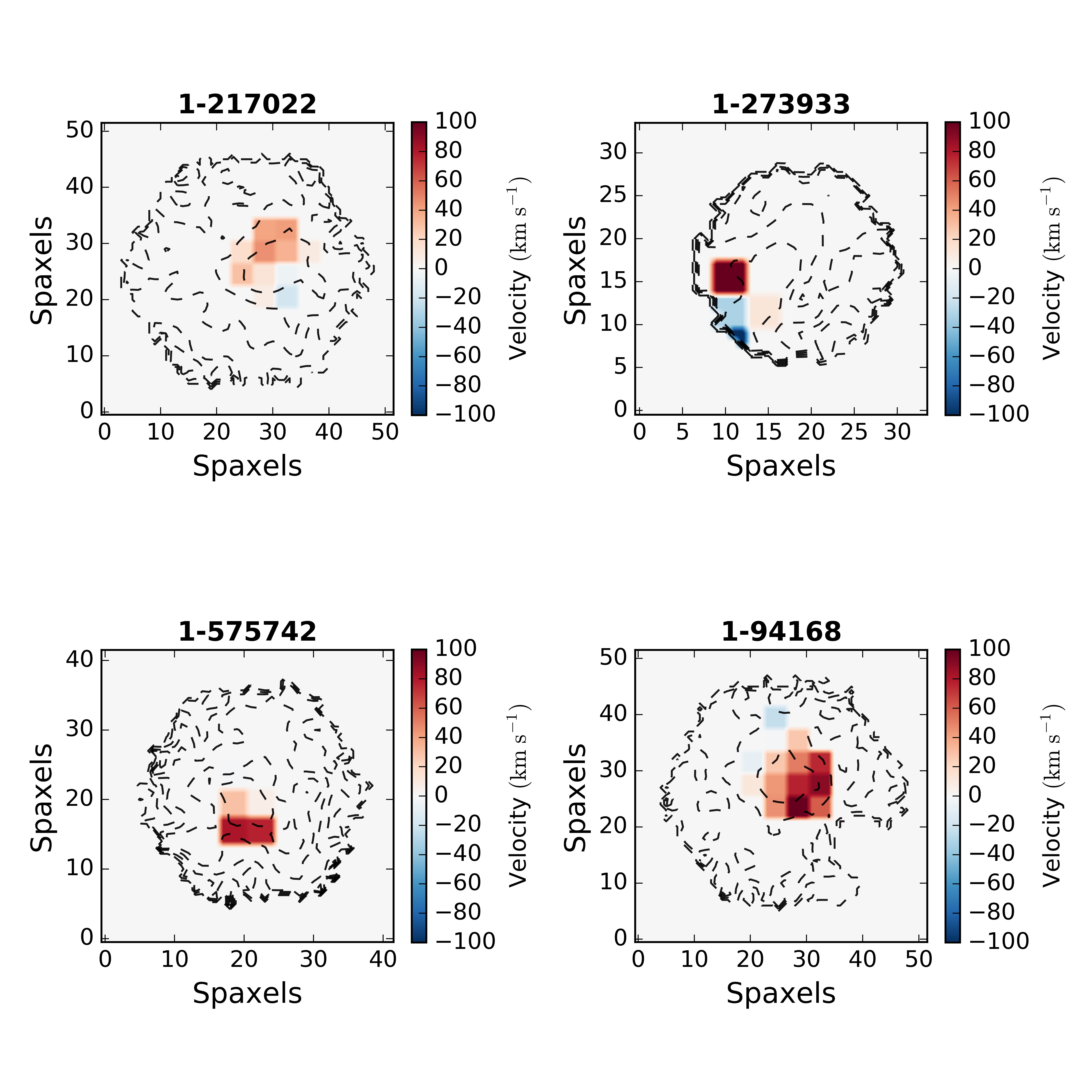}
\caption{  Spatially resolved NaD velocity maps of the same four red geysers as Fig.~\ref{fig:ismmaps}(MaNGA ID: 1-217022, 1-273933, 1-575742 and 1-94168). The velocities are extracted from modeling the NaD absorption spectra with \cite{rupke05} model, wrapped in an MCMC framework. 
\label{fig:velmaps}}
\end{figure}

By a number of measures, the previous sections have illustrated that red geysers, and in particular radio-detected red geysers, feature more prominent levels of ISM-associated cool material as traced by NaD.  We now turn to the kinematics of this material as a way to probe its physical origin and evolution. We first calculate the spatially-resolved kinematics of the cool neutral ISM in red geyser and control samples by measuring the doppler shift of the ISM component of the NaD absorption line with respect to the systemic velocity for each spaxel in every galaxy (see \S\ref{meanew}).

We calculate the mean doppler velocity offset within each galaxy after weighting by the column density, similar to the computation of mean EW. The comparison between the mean velocity in radio detected and not-detected red geyser with control samples (split by radio detection) is shown in Fig.~\ref{fig:intvel}. The radio-RG sample shows an average positive (or redshifted) velocity V$= \rm  41\pm 14~km~s^{-1}$, with respect to the systemic velocity. The nonradio-RG sample also shows a positive velocity, though of lesser amplitude, with a mean value $\rm 13\pm9~km~s^{-1}$. The radio detected and non detected control galaxies, on the other hand, has average V$=\rm 8\pm 12~km~s^{-1} $ and $\rm 4\pm 11~km~s^{-1}$ respectively. 
Thus the radio red geyser sample shows more redshifted velocity on average. 
Looking at the statistics across the samples, we find that $\rm 86\%$ of radio RG sample has V$\geq 0$, while only $\rm 24\%$ and $13\%$ of radio-CS and nonradio-CS galaxies show a redshifted velocity. The radio RG sample show a statistically different distribution from radio- detected and non-radio detected control samples from KS test with p value$=0.009$ and $\rm 6\times 10^{-4}$ respectively. The nonradio-RG however show similar distribution with radio-CS with p$>0.01$, but significant difference with nonradio-CS (p value $= 1\times 10^{-3}$. 

Fig.~\ref{fig:velmaps} shows spatially resolved velocity maps of the NaD-ISM absorption spectra of the same four red geysers as Fig.~\ref{fig:ismmaps} (MaNGA ID: 1-217022, 1-273933, 1-575742, 1-94168). Except for the galaxy in the lower left panel, the rest are radio-detected. The radio detected red geysers show significant redshifted spaxels, generally clustered on one side of the galaxy.  

\subsection{Global stacked kinematics }  \label{int_kin}

\begin{figure}[h!!!] 
\centering
\graphicspath{{./plots/}}
\includegraphics[width = 0.5\textwidth]{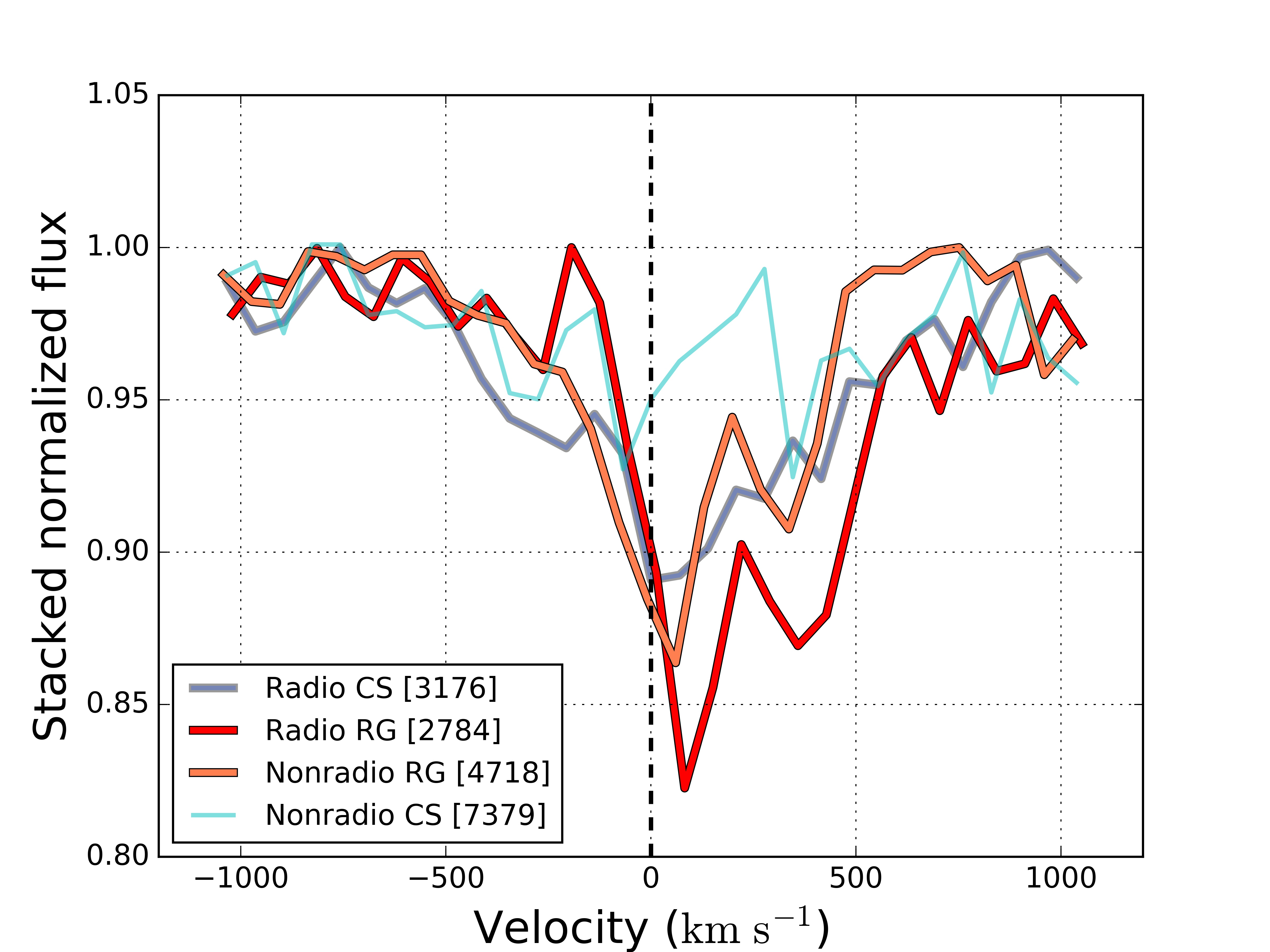}
\caption{ Stacked and normalized NaD spectra containing only the ISM component for radio detected and not-detected red geysers (red and orange) and control sample (blue and cyan) galaxies respectively. The total number of spaxels with detected NaD ISM used in the stacking for each of the galaxy sample is mentioned in the legend in square brackets}. The radio detected red geyser sample show the highest positive velocity (or inflow) relative to the systemic velocity, closely followed by the non radio geysers. The control sample do not exhibit any visible redshift signature in the stacked sample. 
\label{fig:stack}
\end{figure}


We have found a clear, redshifted signature (positive velocity, $\rm V = 41~km~s^{-1}$) in the average of the spatially-resolved NaD offset velocity in the radio detected red geyser galaxies.
However, to estimate these average velocities, we performed Bayesian fitting assuming a specific absorption model to each spaxel (\S\ref{fitting}) and computed the mean velocities after weighing by column density. We have shown in Fig.~\ref{fig:sim} that a small but positive bias ($\leq \rm 5~km~s^{-1}$ at $\rm S/N > 10$ per dispersion element) can emerge from the fitting algorithm used for the kinematic analyses. Thus to ensure that the velocities we get do not arise due to artifacts of model fitting or averaging, we compute the kinematics of stacked NaD profiles from all four sample of galaxies to check if we see similar trends.

We only use spaxels that meet the criteria described in \S \ref{fitting}. Additionally we impose another constraint that the considered spaxels should have a measured EW of the total NaD feature (including both stellar and ISM contribution) to be $>$ 2\AA. This picks out the spaxels with an appreciable NaD signal in each galaxy and provides confidence in the measured equivalent width from the spectra. We then perform a spaxel wise stacking of the continuum normalised spectra around the NaD feature and over each sample. The stacked spectra for all four samples are shown in Fig.~\ref{fig:stack}. The red geyser samples feature a positive velocity offset. The radio-detected red geysers show a higher redshifted velocity ($\rm +53\pm 6~km~s^{-1}$) compared to non-radio-detected geysers ($ 31 \pm 5~\rm km~s^{-1}$). The control samples do not show any significant redshift signature relative to the systemic velocity and show values within $\rm \pm 10~km~s^{-1}$ with large errors (See Table~\ref{tab:summary}, column 5). The errors quoted in Table~\ref{tab:summary} indicate mean 1-$\sigma$ uncertainty over the whole sample. The stacked velocity values agree well with the average resolved kinematic measurements obtained in \S\ref{resolved_kin}.

\section{Discussion}  \label{discussion}

\begin{table*}
	\centering
	\caption{Summary of the NaD-ISM properties obtained for four galaxy samples }
	\label{tab:summary}
	\begin{tabular}{lcccr} 
		\hline
		Sample & Mean EW [\AA] & Fractional area on-sky & Mean resolved V [$\rm km~s^{-1}$] & Stacked V [$\rm km~s^{-1}$]   \\
		\hline
		Radio-red geysers & $0.90^{+0.31}_{-0.30}$ & $0.28^{+0.09}_{-0.07}$ & 40.86 $\pm$ 14.07 & 52.8 $\pm$ 6.27 \\
		Nonradio-red geysers & $0.51^{+0.35}_{-0.33}$ & $0.22^{+0.15}_{-0.13}$ & 13.56 $\pm$ 9.21 & 31.0 $\pm$ 5.02 \\
		Radio control sample & $0.46^{+0.29}_{-0.28}$ & $0.09^{+0.15}_{-0.09}$ & 8.13 $\pm$ 12.10 & 10.2 $\pm$ 7.78\\
		Nonradio control sample & $0.27^{0.41}_{-0.26}$ & $0.1^{+0.1}_{-0.09}$ & 4.00 $\pm$ 11.97 & -28.1 $\pm$ 25.3\\
		\hline
	\end{tabular}
\end{table*}


We have examined the cool ISM properties as probed by NaD in four samples of interest: 30 radio-detected red geysers, 110 non-radio-detected red geysers, 65 radio-detected control sample galaxies and 393 non-radio-detected control sample galaxies. Table.~\ref{tab:summary} summarizes the main findings along with mean 1-$\sigma$ error values calculated within each galaxy sample.  

We find that the radio detected red geyser sample show the highest mean EW (0.9 \AA) from the NaD ISM component, compared to the rest of the samples (Table.~\ref{tab:summary}, column 2). Fig.~\ref{fig:meanew} shows that the average EW in radio-RG sample is about 0.5~\AA ~higher than the control galaxies, irrespective of radio detection. We also find that $ 78\%$ of radio-RG sample possess an appreciable amount of cool gas detected over a projected area $> \rm  5\times 5~kpc^2$, which is the largest compared to the other samples. The NaD in the radio detected red geysers also occupy a larger on-sky area on average, roughly $ 28\%$ of the projected area of the galaxy (Fig.~\ref{fig:areaonsky}). These results imply that the red geyser galaxies, which host low luminosity radio AGNs, typically have an abundant supply of cool gas. A rough estimate of the mass of this detected cool gas is given in the later part of this discussion in \S\ref{sec:coolgas_mass}.



Fig.~\ref{fig:intvel} shows that the average doppler shift of NaD obtained from the spatially resolved kinematics in the radio RG sample is roughly $\rm 40 ~km~s^{-1}$ with respect to the systemic velocity. The non-radio RG show a smaller positive velocity $ \rm 13~km~s^{-1}$. The control galaxies, irrespective of radio detection, show an almost zero velocity within the uncertainties (Table.~\ref{tab:summary}). This result very closely matches with the global stacked kinematics obtained by stacking all the galaxies in each of the four samples. We find that the radio detected red geysers show a strong redshift or positive velocity, of about $\rm 53~km~s^{-1}$ compared to the other samples (Fig.~\ref{fig:stack}, also see Table.~\ref{tab:summary}). Redshift in absorption indicates gas inflowing towards the galaxy, away from the observer. The correlation with radio properties together with inflowing kinematics possibly indicate that the cool gas is fueling the central AGN in the radio detected red geyser galaxies, very similar to the prototypical red geyser in \cite{cheung16}. It is interesting to note that the radio-detected control galaxy sample also displays a moderate amount of ISM-associated cool gas via NaD absorption (mean EW $=$ 0.46 \AA) but shows no inflowing signature as in the red geysers. Our interpretation in alignment with \cite{roy18} is that these radio detected control galaxies, which also host low-luminosity radio AGNs, are not capable of driving energetic winds like the red geysers in present time but they might in the future. This detected cool gas may slowly accumulate over time and can eventually fuel the central AGN which can then trigger the red geyser winds.  

\subsection{Cool gas mass and inflow rates} \label{sec:coolgas_mass}

 In order to get a rough estimate for the implied mass inflow rate in red geysers, we adopt the following equation from \citet{borsani18}:

\begin{equation}\label{eqn:outflowrate}
  \rm  \dot{M}_{out} = \Omega~\mu~m_{H}~N(H)~v~r
\end{equation}

\noindent where $\Omega$ is the solid angle subtended by the wind at its origin, m$_H$ is the mean atomic
weight (with a $\mu$=1.4 correction for relative He abundance), N(H)
is the column density of Hydrogen along the line of sight, v is the doppler velocity, and r is the extent of the NaD absorption in the galaxy with respect to the center.

Making the same assumptions as \citet{rupke05}, the column density of hydrogen can be written as:

\begin{equation}
  \rm  N(H) = \frac{N(Na\ I)}{\chi(Na\  I)~d(Na\  I)~Z(Na\  I)}
\end{equation}

where  N(Na I) is the sodium column density,
$\chi$(Na I)= N(Na I)/N(Na) is the assumed ionization fraction,
d(Na I) is the fraction depletion onto dust, and Z(Na I) is the Na abundance. We assume a 90$\%$ ionization fraction ($\chi$(Na I)=0.1), a Galactic value for the depletion onto dust (log d(Na I)=-0.95), and 
solar metallicity (Z$\rm_{solar}$(Na I)=log[N(Na)/N(H)]=-5.69) similar to that assumed in \citet{rupke05}. Putting the log N(Na I) [$\rm cm^{-2}$] values which range from 11.6 to 13.7 obtained from modeling the NaD absorption feature in radio detected red geyser sample, we obtain an average log N(H)/$\rm cm^{-2}~\sim 19.50-21.30$. This is roughly in agreement with the log N(H)/$\rm cm^{-2}~\sim 21$ estimated according to \cite{bohlin} from the average dust extinction in similar galaxies.

With these assumptions and setting $\Omega v r$ in Eq.\ref{eqn:outflowrate} equal to $C_f A_{\rm NaD} / t_{acc}$, where $\rm A_{NaD}$ is the on-sky projected area of the galaxy comprising NaD absorption from \S\ref{sec:onsky_area} and $\rm t_{acc}$ is the accretion timescale $\sim \rm \frac{R_e}{ v }$ , we get:

\begin{equation} \label{eqn:mout}
\begin{aligned}
    \rm \dot{M}_{out} = \mu~m_{H}~C_f~N(H)~A_{NaD}~\frac{v}{R_e}
\end{aligned}
\end{equation}

The total mass of the gas can be estimated by the assumption: $\rm M_{tot} = \dot{M}_{out} \times t_{acc} $. 

\begin{equation} \label{eqn:mtot}
    \rm M_{tot} = \mu~m_{H}~C_f~N(H)~A_{NaD}
\end{equation}

The cloud covering factor C$\rm _{f}$ is taken to be 0.4, the average value obtained from our model fits. 
For the radio-detected red geysers, the NaD velocity (v) varies from $\rm 10~km~s^{-1} - 100~km~s^{-1}$ (Figure.~\ref{fig:intvel}) with a mean of $\sim\rm 45~km~s^{-1}$ (Table.~\ref{tab:summary}, column 4 and 5). The estimated total mass ranges from $\rm M_{cool} \sim 0.1 \times 10^{7}~M_{\odot} - 5. \times 10^{8}~M_{\odot}$ within the radio-detected red geyser sample while the mass inflow rate  spans a large range ~$\sim \rm 0.02 - 5~M_{\odot}~yr^{-1}$. The estimated hydrogen column density, mass flow rates and the mass of the gas for each galaxy in the radio RG sample is given in Table.~\ref{tab:inflow} along with their 1-$\sigma$ uncertainties. The 1-$\sigma$ errors in the column density are derived from MCMC fitting of the absorption line model (\S\ref{fitting}) to the observed NaD spectrum in each spaxel, satisfying the detection criteria, within a galaxy and adding them in quadrature. The corresponding errors in flow rate and gas mass are obtained from simple error propagation.

A few galaxies show negative velocity (or blueshift) indicating possible outflowing gas and the outflow rate, estimated from the same eqn.~\ref{eqn:mout}, is given with a negative sign in the table. As expected, the majority of radio-detected red geysers show inflowing gas. Those galaxies with no NaD ISM component detected in any spaxels (following \S\ref{fitting}) are tabulated as having zero inflow rate and $\rm M_{tot}$ although there may be cool gas present below our detection level.

\subsection{Origin of inflowing gas}

Infalling gas in red geysers at the estimated accretion rates can come from a variety of sources: 

\begin{itemize}[noitemsep]

\item Accretion from outside the galaxy and its halo.  Galaxy mergers, specifically minor mergers, present an obvious mechanism \citep{sanders88, weston17, kaviraj14, pace14,
martin18}. 
\item A completely internal source, such as gas injected into the interstellar medium during stellar mass loss \citep{conroy15} which undergoes cooling and condensation, perhaps eventually feeding an AGN \citep{ciotti97, ciotti01, ciotti07}. 
\item Fountain scenario where the warm ionized outflowing gas, driven out by the central AGN from earlier geyser episodes, rises into the halo, cools and condenses as it expands, and then falls back onto the galaxy \citep[e.g., ][]{shapiro76, bregman80, norman89,tremblay16, tremblay18, voit17}.

\end{itemize}

Numerous observations have measured the merger rates in nearby galaxies \citep[see][and references therein]{ellison15} and several authors have emphasized widespread merger activity when observational evidence sensitive to long timescales and low mass ratios is scrutinized \citep[e.g.][]{dokkum05, sancisi08}.  
For galaxies with stellar masses between $\rm 10^{10.5}-10^{12}~M_{\odot}$, the average predicted merger rate per galaxy (assuming typical mass ratio of minor mergers $\sim \rm 1:10$) from semi-empirical models and hydrodynamical simulations is roughly $\sim \rm 0.1~Gyr^{-1}$
 \citep{hopkins10, oleary21}. Thus galaxy mergers with a 1:10 (or greater) mass ratio can generate an $M_*$ accretion rate of $\rm \frac{0.1 \times 0.1 \times 1.\times 10^{11}~M_{\odot}}{10^{9}~yr} = 1~M_{\odot}~yr^{-1}$ on average. If 10$\%$ of this value lies in cold gas, the estimated accretion rate is roughly $\rm 0.1~ M_{\odot}~yr^{-1}$. 
 This is similar to the accretion rates quoted in \cite{capelo15} which performed detailed simulations by exploring a wide merger parameter space including different initial mass ratios, orbital configurations and gas fractions with high spatial and temporal resolution. 
 The average cool gas inflow rate we estimate from NaD in red geysers is similar to the accretion rate implied by minor mergers.  Merging gas-rich dwarfs may therefore be an important sources of the infalling gas.  The apparently higher detection rate of NaD-associated inflows in red geysers compared to control sample galaxies may be telling us that red geysers are a signpost of accretion activity.  Indeed, some red geysers in our sample with accretion rate $\rm \sim 5~M_{\odot}~yr^{-1}$ may be signaling a peak in the instantaneous accretion rate during a merger event.

In addition to mergers, \cite{ciotti91, ciotti07} predicted that a significant amount of gas, $\sim \rm 20 - 30 \%$ of the initial stellar mass of the galaxy, is donated throughout the galaxy over the course of stellar evolution and supernova explosions.  This gas can be accreted onto an AGN. The radiative cooling time of this warm gas in typical elliptical galaxies is roughly $\rm \sim 10^{7.5-8} yr$. \cite{ciotti07} quoted a mass return rate of $\sim \rm 5 - 10~M_{\odot}~yr^{-1}$ from the stellar populations over the cooling timescale, which exceeds the cool gas inflow rates we observe in red geysers. 
Moreover, some fraction of warm ionized gas ($\rm M_{warm} \sim 10^{5-6}~M_{\odot}$), ejected by the red geyser winds, may also accumulate within the ISM or the CGM of the galaxies over time. After cooling and condensation, this material might also  become a source for the observed inflows. 

Considering the different sources of infalling gas discussed above, 
we have found from visual inspection that most of radio-detected red geysers show a spatially irregular and clumpy NaD distribution (see Fig.~\ref{fig:ismmaps}). 
Additionally, we see from Fig \ref{fig:areaonsky} that the average fractional area occupied by the NaD absorption is around $\sim \rm 30\%$ of the projected area of the entire galaxy for the radio red geyser sample, which implies a concentrated, lumpy distribution of gas confined in a small region. The recycled gas from the fountain scenario might be expected to present a more diffuse and smooth component than is observed here.

Cooling ISM gas originating from stellar winds and supernovae might also be expected to cover a larger area of the whole galaxy.
Moreover, the cool gas originating from stars might show greater correlations with the stellar kinematics which is not observed here. 

It is also interesting to consider the relatively short timescales of dust-enshrouded cool gas.
As discussed in \cite{rowlands12}, dust in early type galaxies get destroyed on relatively short timescales, ranging from 50 - 500 Myr, due to thermal sputtering from the hot X-ray halo surrounding the ISM or from shocks generated from Type 1a supernova.  The signatures of clumpy, dust-associated NaD gas are likely short-lived, suggesting stochastic inflow episodes from merging systems over a more continuous `rain' of recycled material.

Finally, we note that the detected cool gas in red geysers, with masses ranging from $\rm 10^7 - 10^8~ M_{\odot}$ and average accretion rates  $\sim 0.1-\rm 5~M_{\odot}~yr^{-1}$, might be expected to trigger a star formation rate of at least $\rm 1~ M_{\odot}~yr^{-1}$. However, the observed average star formation rate ($\rm \sim 10^{-2}~M_{\odot}~yr^{-1}$) is
several orders of magnitude lower, as evident from Fig.~\ref{fig:sfr}. This again is suggestive of active feedback mechanisms, perhaps related to the red geyser winds themselves, that are effective in suppressing star formation in these galaxies.

\begin{table*}
	\centering
	\caption{Estimated mass and inflow rate of cool gas for the radio detected red geyser galaxies }
	\label{tab:inflow}
	\begin{tabular}{|l|c|c|c|r|} 
		\hline
	\textbf{MaNGA-ID} & \textbf{Radio luminosity L} [$\rm 10^{22}~W~Hz^{-1}$] & \textbf{Hydrogen Column Density log N(H)} [$\rm cm^{-2}$] & \textbf{Inflow rate} [$\rm M_{\odot}~yr^{-1}$] & \textbf{Gas mass} [$\rm 10^{8}~M_{\odot}$]   \\
		\hline\hline
		1-352569 & 8.79 & 20.45$\pm$1.51 & 0.19$\pm$0.02 & 2.4$\pm$0.21  \\
        1-23958 &  0.32 & 20.89$\pm$0.45 & 0.47$\pm$0.03 & 1.3$\pm$0.10\\
        1-113668 & 5.92 & 20.38$\pm$2.12 & $-$0.41$\pm 0.04$  & 3.5 $\pm 0.36$ \\
        1-595166 & 3.19 & 20.52$\pm$1.99 & $-$0.78$\pm 0.04$ & 2.5 $\pm 0.13$\\
        1-550578 & 7.81 & 20.28$\pm 1.73$ & $-$0.23$\pm 0.01$ & 0.96 $\pm 0.10$\\
        1-634825 & 1.26 & 21.20$\pm0.28$ & 1.71$\pm 0.22$ & 4.3 $\pm 0.26$\\
        1-37036 &  0.12 & 20.03$\pm 2.07$ & 0.02$\pm 2\times 10^{-3}$ & 0.03 $\pm 5 \times 10^{-3}$\\
        1-43718 & 1.15 & 20.32$\pm 1.20$ & 0.05$\pm 9\times 10^{-3}$ & 0.13 $\pm 0.02$\\
        1-218116 & 2.06 & 0.0 & 0.0 & 0.0\\
        1-217022 & 0.16 & 21.27$\pm 0.14$ & 0.46$\pm 0.12$ & 2.4 $\pm 0.18$\\
        1-256446 & 2.29 & 20.81$\pm 0.60$ & 1.80$\pm 0.10$ & 1.9 $\pm 0.22$\\
        1-94168 & 0.52 & 21.11$\pm 0.27$ & 1.22$\pm 0.08$ & 3.7 $\pm 0.24$\\
        1-209926 & 3.36 & 20.68$\pm 0.24$ & 0.78$\pm 0.07$ & 5.1 $\pm 0.48$\\
        1-273933 & 2.46 & 20.47$\pm 0.96$ & $-$0.03$\pm 0.02$ & 0.3 $\pm 0.04$\\
        1-245451 & 5.78 & 21.07$\pm 0.14$ & 5.67$\pm 0.22$ & 5.9 $\pm 0.62$\\
        1-198182 & 1.30 & 0.0 & 0.0 & 0.0 \\
        1-210863 & 0.95 & 19.48$\pm 5.46$ & 0.01$\pm 5\times 10^{-3}$ & 0.01 $\pm 3 \times 10^{-3} $ \\
        1-289864 & 77.52 & 0.0 & 0.0 & 0.0\\
        1-24104 & 0.36 & 20.53$\pm 1.54$ & 0.38$\pm 0.02$ & 0.7 $\pm 0.02$\\
        1-378770 & 45.32 & 20.49$\pm 0.51$ & 0.88$\pm 0.06$ & 3.3 $\pm 0.42$\\
        1-279073 & 0.54 & 20.04$\pm 1.05$ & 0.01$\pm 3\times 10^{-3}$ & 0.02 $\pm 5\times 10^{-3}$\\
        1-188530 & 8.37 &  20.37$\pm 0.93$ & 3.21$\pm 0.2$ & 3.9 $\pm 0.05$\\
        1-209772 & 23.84 & 20.10$\pm 3.99$ & 2.8$\pm 0.5$ & 3.2 $\pm 0.66$\\ 
        1-268789 & 6.51 & 20.19$\pm 0.81$ & 0.04$\pm 0.01$ & 1.0 $\pm 0.02$\\
        1-567948 & 0.13 & 19.90$\pm 2.32$ & 0.4$\pm 0.1$ & 2.1 $\pm 0.3$\\
        1-96290 & 6.22 & 20.27$\pm 1.50$ & 0.27$\pm 0.03$ & 2.0 $\pm 0.24$\\
        1-37440 & 0.06 & 20.9$\pm 0.55$ & 2.38$\pm 0.01$ & 1.1 $\pm 0.04$\\
        1-321221 & 0.25 & 0.0 & 0.0 & 0.0 \\
        1-322336 & 5.64 & 20.62$\pm 0.31$ & 0.35$\pm 0.06$ & 2.3 $\pm 0.40$\\
        1-627331 & 0.91 & 0.0 & 0.0 & 0.0\\
		\hline
	\end{tabular}
\end{table*}

\section{Conclusion}  \label{conclusion}

We have performed a set of analyses of the sodium doublet absorption feature as a tracer of cool ISM gas in red geyser galaxies and a matched control sample using spatially-resolved data from the SDSS IV-MaNGA survey. After carefully subtracting the stellar contribution, we have found an excess in  NaD absorption which we ascribe to the presence of cool material in the ISM.  We study the properties of this gas in our radio-detected red geyser sample (30 galaxies) and compare to several other samples - namely, 110 non-radio detected red geysers, 65 radio-detected and 393 non radio detected control samples.

We measure the kinematic behavior of the cool gas by computing the average velocity offsets in spatially resolved maps, integrated offsets per galaxy, and stacked kinematics of ISM NaD spectra across all four galaxy samples. We find that the NaD in the radio-detected red geysers shows a clear redshift ($ 40-50 \rm~km~s^{-1})$ with respect to the systemic velocity.  We interpret this result as an indication that the cool gas is inflowing into the galaxy center. The non-radio-detected sample of red geysers also show a similar redshift, but of slightly smaller amplitude on average
($\sim 15 - 30 \rm~km~s^{-1}$). The control samples however show average velocity offsets around zero or slightly negative ($\rm < 10~km~s^{-1}$). 

In order to check that the observed redshift is robust due to the systematics of the fitting technique itself, we tested our fitting approach using simulated spectra of varying signal-to-noise and different input velocities (Fig.~\ref{fig:sim}). We find that at the signal-to-noise of our observation ($>10$ per dispersion element), there exist a modest systematic bias of $\leq \rm 5~km~s^{-1}$ which is, however, much smaller than the velocity inflow values we observe.

The accretion of cool gas inflowing into the center coupled with the radio enhancement in the radio red geyser sample indicates that the central low luminosity radio AGN is possibly fueled by the cool neutral gas clouds in the ISM. Both external sources (like mergers) and internal processes (stellar mass loss, accumulation of gas from relic geyser winds) can explain the observed accretion rates, which span a vast range $\sim \rm 0.02 - 5~M_{\odot}~yr^{-1}$ with a estimated total mass of the cool gas to be $\sim \rm 10^{7}-10^{8}~M_{\odot}$. 
Although this gas is capable of triggering an easily detectable level of star formation ($\sim \rm 1~M_{\odot}~yr^{-1}$), the observed quiescent nature of the red geysers with a SFR almost 2 orders of magnitude lower indicates evidence of maintenance-mode feedback in action and possibly associated with red geyser phenomenon itself.



\section*{Acknowledgement}
The authors thank the anonymous referee for helpful comments and suggestions that significantly improved the mansucript. 
This research was supported by the National Science Foundation under Award No. 1816388. 
RR thanks the brazilian funding agencies CNPq, CAPES and FAPERGS.
Funding for the Sloan Digital Sky Survey IV has been provided by the Alfred P. Sloan Foundation, the U.S. Department of Energy Office of Science, and the Participating Institutions. SDSS-IV acknowledges
support and resources from the Center for High-Performance Computing at
the University of Utah. The SDSS web site is \href{http://www.sdss.org}{www.sdss.org}.

SDSS-IV is managed by the Astrophysical Research Consortium for the 
Participating Institutions of the SDSS Collaboration including the 
Brazilian Participation Group, the Carnegie Institution for Science, 
Carnegie Mellon University, the Chilean Participation Group, the French Participation Group, Harvard-Smithsonian Center for Astrophysics, 
Instituto de Astrof\'isica de Canarias, The Johns Hopkins University, 
Kavli Institute for the Physics and Mathematics of the Universe (IPMU) / 
University of Tokyo, the Korean Participation Group, Lawrence Berkeley National Laboratory, 
Leibniz Institut f\"ur Astrophysik Potsdam (AIP),  
Max-Planck-Institut f\"ur Astronomie (MPIA Heidelberg), 
Max-Planck-Institut f\"ur Astrophysik (MPA Garching), 
Max-Planck-Institut f\"ur Extraterrestrische Physik (MPE), 
National Astronomical Observatories of China, New Mexico State University, 
New York University, University of Notre Dame, 
Observat\'ario Nacional / MCTI, The Ohio State University, 
Pennsylvania State University, Shanghai Astronomical Observatory, 
United Kingdom Participation Group,
Universidad Nacional Aut\'onoma de M\'exico, University of Arizona, 
University of Colorado Boulder, University of Oxford, University of Portsmouth, 
University of Utah, University of Virginia, University of Washington, University of Wisconsin, 
Vanderbilt University, and Yale University.\\


\end{document}